\documentclass[%
reprint,
 unsortedaddress,
showkeys,
 amsmath,amssymb,
 aps,prb
]{revtex4-2}

\usepackage{amsmath,amssymb,ulem}
\usepackage{euscript} % for \EuScript{}
\usepackage{newtxtext,newtxmath}% Use Times as default font
\usepackage{graphicx}% Include figure files
\usepackage{dcolumn}% Align table columns on decimal point
\usepackage{bm}% bold greek math
\usepackage{braket}% bra ket notation
\usepackage[usenames, dvipsnames]{color}
\usepackage{float}
\usepackage{hyperref}% add hypertext capabilities
\begin{document}

\title{Synthetic Kramers pair in phononic elastic plates and\\
helical edge states on a dislocation interface}
\author{Ting-Wei Liu}
\author{Fabio Semperlotti}
 \email{fsemperl@purdue.edu}
\affiliation{Ray W. Herrick Laboratories, School of Mechanical Engineering, Purdue University, West Lafayette, Indiana 47907, USA.}
\begin{abstract}
In conventional theories, topological band properties are intrinsic characteristics of the bulk material and do not depend on the choice of the reference frame. In this scenario, the principle of bulk-edge correspondence can be used to predict the existence of edge states between topologically distinct materials. In this study, we propose and carefully examine a 2D elastic phononic plate with a Kekul\'e-distorted honeycomb pattern engraved on it. It is found that the pseudospin and the pseudospin-dependent Chern numbers are not invariant properties, and the $\mathbb{Z}_2$ number is no longer a sufficient indicator to examine the existence of the edge state. The distinctive pseudospin texture and the pseudomagnetic field are also revealed.
Finally, we successfully devise and experimentally implement the synthetic helical edge states on a dislocation interface connecting two subdomains with bulk pattern identical up to a relative translation. The edge state is also imaged via laser vibrometry.
\end{abstract}


\maketitle

The study of topological states of matter has rapidly grown over the past few decades \cite{ReviewKaneTI,ReviewNiu,ReviewZhang}. While the field of topological materials originated and had particular significance for the broader area of quantum mechanics, in recent years these works served as an inspiration to explore the existence of analog topological effects in classical waveguide systems \cite{ReviewTopologicalPhotonics,ReviewTopologicalPhotonics2,ReviewTopologicalPhotonics3,reviewtopologicalphononics2018,reviewtopologicalphononics2019}. On one side, these analog quantum mechanical mechanisms offered engineers a powerful route to design waveguides (either photonic, acoustic, or elastic) immune to backscattering generated by disorder or defects. On the other side, these same mechanisms were seen by physicists as ideal platforms to explore the effect of design parameters and to experimentally validate important fundamental concepts without the significant complexities imposed by the quantum scales. In other terms, the artificial photonic and phononic structures were treated as surrogate platforms to understand and develop the topological band theory and its practical implications.
%while physicists find the artificial lattice a new platform to investigate the topological band properties which is much easier to manipulate and observe contrasted to the limited degree of freedom and extreme condition required to study the electronic system.

To-date, several analog mechanisms have been explored and validated for classical waveguides. Initial attempts focused on the implementation and transposition of the concept of quantum Hall effect. However, these systems required the breaking of time reversal symmetry (TRS), which imposes significant practical complexities due to the need for either special magneto-optic and elastic materials, or for carefully controlled external input \cite{Haldane,zhang2005experimental,photonicexternal2008,photonicexternal2008_2,photonicexternal2018,TopoGyroExp,TunableTopoPnC-Flow,TopoSound-Flow,TopoPhon-Gyro,TopologicalAcoustics-Flow}. 
More recently, mechanisms analog to TRS-preserved quantum spin Hall effect (QSHE) \cite{KaneMele,bhz,photonicgapped2015,AcousticTIAir,fold2017,zonefold2018,zonefold2,kekuleprl2017,photonicgapped2018,gapped2018,kekulenjp2018,kekule2019,deng2019comparison,snowflake,AcousticTIPlate,miniaci2018experimental} and quantum valley Hall effect \cite{fewlayer,bilayerDWexp,bilayerDW,ValleyContrasting,ValleySonicEdge,pal2017edge,vila2017observation,photonicvalley2018,myqv,diatom,deng2019comparison,QVEXP,QVABH}
were also proposed.
These systems could be built based on ordinary dielectric or linearly elastic materials, and only required the breaking of spatial symmetry, which was a considerably more practical approach.
    Other studies proposed the use of negative indices, instead of artificial lattice structures, to realize analogues to the QSHE in electromagnetic and acoustic systems \cite{bliokh2019topological,bliokh2015quantum,bliokh2015spin,bliokh2019klein,leykam2020edge,bliokh2019transverse}.
Among the TRS-preserved mechanisms, bosonic systems can leverage the same valley degree of freedom to realize an analog valley Hall effect.
While the valley Hall effect and the associated backscattering immunity of the edge states count on the large separation of states in momentum space, the classical analog to QSH systems relies on the decoupling of synthetic ``pseudospin'' states. These states can potentially yield counter-propagating edge states that are close in momentum space while still being robust against backscattering.

Despite these successful initial implementations, the emulation of a quantum spin Hall topological insulator has proven to be a remarkably more complicated task.
It is well known that the $\pm \frac{1}{2}$-spin electronic systems exhibit so-called Kramers pairs (given that the time-reversal operator $\mathcal{T}$ applied twice returns the initial state with additional negative sign, that is $\mathcal{T}^2=-\mathbf{1}$, where $\mathbf{1}$ denotes the identity operator). It follows that under TRS, every eigenstate has a partner with opposite momentum and spin and the same energy spectrum, that is $E(\mathbf{k},\uparrow)=E(-\mathbf{k},\downarrow)$. This condition guarantees the existence of gapless helical edge states that are time reversal counterparts of each other with opposite spins. While both photons and phonons are bosonic in nature, they do not possess the same intrinsic attributes. Researchers have proposed different ideas to synthesize properties analog to electron spins, often referred to as ``pseudospins''. In phononic elastic systems, examples include mixing of symmetric and antisymmetric Lamb modes in waveguides \cite{AcousticTIPlate,miniaci2018experimental}, and a ``zone-folding'' method \cite{photonicgapped2015,AcousticTIAir,fold2017,zonefold2018,zonefold2,kekuleprl2017,photonicgapped2018,gapped2018,kekulenjp2018,kekule2019,deng2019comparison,snowflake} to attain a doubly degenerate Dirac cone at the center of the Brillouin zone that acts as a degree of freedom that resembles electron spins.
Nevertheless, it was reported that phononic systems exploiting this mechanism could give rise to gapped edge states at zero momentum
where the $\omega$-$\mathbf{k}$ dispersion curves of the counter-propagating edge states repel each other, due to coupling between them
%when the two edge states are close in the frequency ($\omega$)-momentum ($\mathbf{k}$) spectra 
\cite{gapped2018,kekuleprl2017,photonicgapped2018,gapped2018,kekulenjp2018,kekule2019,photonicgapped2015,AcousticTIAir,fold2017,zonefold2018,zonefold2,deng2019comparison}. These results showed that the edge states are not a Kramers pair and do not have a continuous spectrum across the bulk band gap. In addition, although the zone-folding approach is already widely adopted, previous studies concentrated on mapping the system back to the electronic counterpart but usually omitted explaining some discrepancies between the synthetic phononic pseudospins and the electron's intrinsic spin; hence, leaving behind some obscure points such as the indeterminate pseudospin states, and the seemingly indistinguishable topological phases.

In this paper, we propose an elastic analog of a topological insulator based on a 2D phononic waveguide designed according to a Kekul\'e distorted honeycomb pattern. The study uncovers interesting anomalous properties associated with the generation of synthetic pseudospins and peculiar discrepancies with electron spins. These anomalous properties include indeterminate pseudospin-dependent Chern numbers and indistinguishable topological states,
which are characteristic of the ``zone-folding'' approach and different from electronic QSH systems. Although the concepts of pseudospins and zone-folding have been discussed in the literature, a series of questions were left unexplored. The discrepancy between the classical ``zone-folding system'' and electronic QSH systems leads to consider the former as an ``imperfect analogue'' since proper ``helical'' edge state cannot be realized. In other terms, the counter-propagating edge states are found to be gapped and with strong coupling at $\mathbf{k}=0$. However, as we face the discrepancy and investigate its cause, we identify a method to realize truly gapless helical edge states such that the counter-propagating edge states are fully decoupled. In addition, we leverage the unique ambiguity of the topological state to achieve edge states along a \textit{dislocation interface} connecting two sections of the same bulk lattice. This latter condition is not achievable in QSH systems in which edge states exist on an interface between materials exhibiting distinct and determinate topological character.

This concept is also experimentally validated by directly imaging the response of the phononic lattice via laser vibrometry, and by extracting the spectrum of the edge state which matches well with the theoretical and numerical predictions.
It is worth to highlight that these findings are particularly relevant from the practical perspective of a waveguide design. In fact, not only this approach leads to robust and continuous gapless edge states, but it also allows exploiting the concept of dislocation that greatly simplifies the design. It is expected that the current design can be very well suited to achieve vibration and structure-borne noise control capabilities fully integrated in lightweight load bearing structures.

\begin{figure}[ht]
\includegraphics[width=0.4\textwidth]{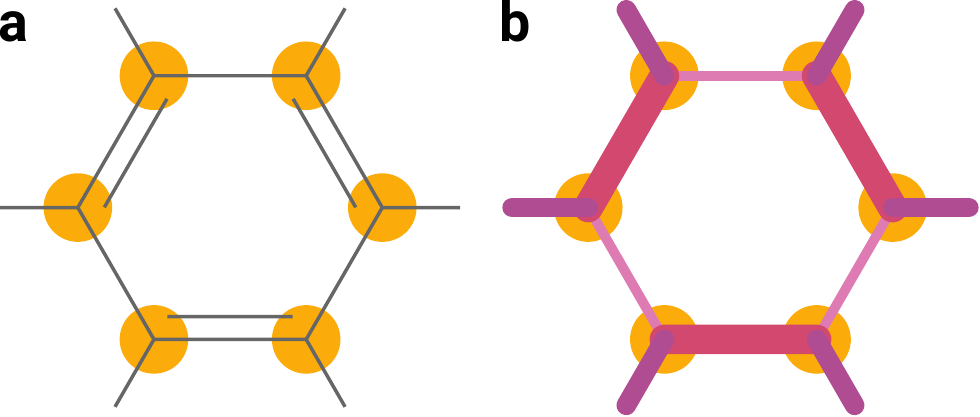}% kekule
\caption{\label{fig:kekule}
(a) The benzene structure suggested by Kekul\'e which contains alternating double and single bonds. (b) The extended Kekul\'e distorted pattern used in our design. The bonds are replaced by the grooves (denoted in different colors) having different and continuously varying depth (illustrated by different line weight).
}
\end{figure}

\begin{figure*}[ht]
\includegraphics[width=0.8\textwidth]{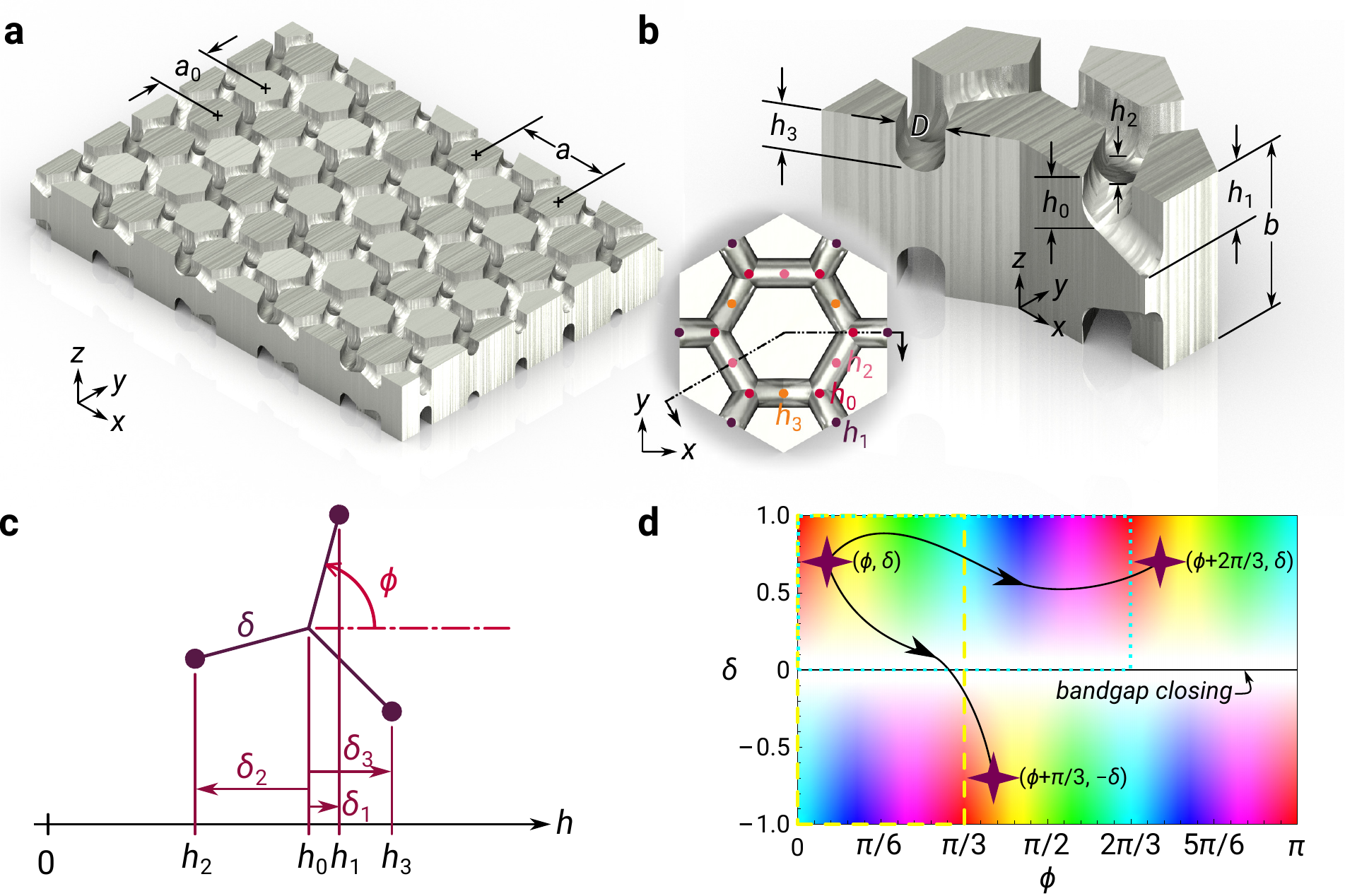}% lattice geometry
\caption{\label{fig:geom}
(a) An illustration of the phononic elastic lattice consisting in an aluminum plate with a honeycomb-like groove pattern symmetrically engraved on both sides. Before the application of the Kekul\'e distortion, the lattice has constant $a_0$. After the distortion is applied, the lattice constant becomes $a=\sqrt{3}a_0$. (b) An isometric sectional view of a Wigner-Seitz unit cell of a distorted lattice centered at a $C_3$ axis. Three different depths $h_{1,2,3}$ are assigned to the midpoints of all the hexagon edges, and the average depth $h_0=(h_1+h_2+h_3)/3$ is assigned at the vertices joining the edges. The grooves are constructed by linearly connecting neighboring grooves. The inset shows the top view of the unit cell. (c) A graphical representation showing the relation between the parameters $(h_0, \delta, \phi)$ and $(h_1, h_2, h_3)$. (d) The parametric space to generate all admissible lattices. The color coding should be interpreted as follows: the hue indicates the parameter $\phi$ mod $2\pi/3$; the color lightness indicates the distortion strength $1-|\delta|$. Different points with the same color yield the same bulk lattice pattern up to a rigid body translation (e.g., the three purple stars). Two examples of the irreducible parametric domains to generate a unique bulk lattice pattern are shown as the yellow dashed and the cyan dotted box. Depending on different paths, two distinct bulk lattice can evolve from one to another with or without band gap closing.
}
\end{figure*}

The proposed phononic lattice consists in an aluminum thin waveguide (i.e. a thin plate) with a honeycomb-like groove pattern symmetrically engraved on both sides. The groove has a variable thickness according to an extended Kekul\'e distortion pattern. Fig.~\ref{fig:kekule}~(a) shows the benzene structure, originally suggested by Kekul\'e, which contains alternating double and single bonds. We adopt an extended version of the Kekul\'e distortion to define the groove depth of the honeycomb cell which allows three different ``bonds'' (i.e. the individual grooves forming the side of the cell) that can change their values (i.e. their depths) continuously, as shown in Fig.~\ref{fig:kekule}~(b).
As shown in Fig.~\ref{fig:geom} (a), the phononic plate has thickness $b=1/4$ inch (6.35 mm). Before incorporating the Kekul\'e distortion, the honeycomb groove has a constant depth $h_0$ and a lattice constant $a_0$. The Kekul\'e distortion is then introduced by varying the groove depth according to the prescribed pattern. Three different depths $h_{1,2,3}$ are assigned at the midpoints of the hexagon edges, and the average depth $h_0=(h_1+h_2+h_3)/3$ is assigned at the vertices of the hexagons joining the edges. The grooves are constructed by linearly connecting neighboring grooves, as schematically shown in the isometric view in Fig.~\ref{fig:geom} (b). Figure \ref{fig:geom} (b) also shows a top view of the Wigner-Seitz unit cell in the inset. Note that the Kekul\'e perturbation enlarges the primitive unit cell and the new lattice constant becomes $a=\sqrt{3}a_0$, as shown in Fig.~\ref{fig:geom} (a). Also, the lattice symmetry is degraded from $D_6$ to $D_3$ (or $C_{6v}$ to $C_{3v}$ in the context of 2D lattices).
To facilitate the fabrication process, the groove width was chosen $D=1/16$ inch, dictated by the diameter of the (ball-end) machining tool. Also the groove bottom was rounded so to match the actual slot cut by the tool. Implementing these few considerations allowed a substantial ease of fabrication via a computer numerical control (CNC) mill.

The three characteristic depths $h_{1,2,3}$ can be easily described by a set of parameters $(h_0, \delta, \phi)$
\begin{align} \label{eq:dist}
h_i &= h_0 + \delta_i,\nonumber\\
\delta_i &= \delta \cos\left(\phi + (i-1)\frac{2\pi}{3}\right), \quad i = 1,2,3.
\end{align}
This parameter set $(h_0, \delta, \phi)\in \mathbb{R}^3$ can be considered as the natural coordinate system of the parametric space $(h_1, h_2, h_3)$ governing the band structure. The first parameter, the average depth $h_0$, affects the overall frequency shift of the phononic band structure, and plays no role in the topological transition. The second parameter $\delta$ controls the magnitude of the Kekul\'e distortion and therefore is related to the opening of the band gap. The parameter $\phi$ is the last free degree of freedom in the 3D parameter space and, will be shown later, is at the root of the peculiar behavior of the lattice. A graphical representation plotting the relation between $(h_0, \delta, \phi)$ and $(h_1, h_2, h_3)$ is shown in Fig.~\ref{fig:geom} (c).
Note that given any real-number sequence  $(h_1, h_2, h_3)$, one can always find an equivalent parameter set  $(h_0, \delta, \phi)$ that yields $(h_1, h_2, h_3)$. Given that the depth value must be positive, we can restrict $h_0>0$ and $-1<\delta<1$.

Due to the properties of the cosine function, we have
%$h_i|_{\phi=\phi'+\pi,\delta=\delta'}=h_i|_{\phi=\phi',\delta=-\delta'}$, 
$h_i|_{(\phi+\pi,\delta)}=h_i|_{(\phi,-\delta)}$, 
for all given $(\phi,\delta)$. Thus, the domain of $(\phi,\delta)$ can be restricted to $\left[0,\pi\right) \times (-1,1) $ while remaining capable of generating all depth sequences $(h_1, h_2, h_3)$, as shown in Fig.~\ref{fig:geom} (d). Another interesting aspect is that given any three sequences $(h_1, h_2, h_3)$, $(h_2, h_3, h_1)$ and $(h_3, h_1, h_2)$ composed of the same three numbers with the same cyclic permutation, they yield the same bulk lattice up to some rigid body translation (or rotation) of the lattice (or, equivalently, of the reference frame). For example, the permutation $(h_1, h_2, h_3)\rightarrow (h_3, h_1, h_2)$ is identical to a translation of the entire lattice by a vector $\left(0, -a_0\right)$ (see the inset in Fig.~\ref{fig:geom} (b)), or a counterclockwise rotation of $2\pi/3$ with respect to the point labeled $h_0$ along the $z$-axis in the inset of Fig.~\ref{fig:geom} (b). It follows that the parameter space in Fig.~\ref{fig:geom} (d) contains three copies of a certain subdomain that can generate the same bulk lattice. Two examples of the irreducible subdomain are highlighted by the yellow dashed and the cyan dotted boxes in \ref{fig:geom} (d). One should find that $(\phi,\delta)$, $(\phi+\pi/3,-\delta)$ and $(\phi+2\pi/3,\delta)$ yield the same bulk lattice up to a rigid rotation or translation, for all given $(\phi,\delta)$. Fig.~\ref{fig:geom} (d) shows all the possible bulk lattice patterns identified by colors.
Specifically, the color reads $({\rm hue},{\rm lightness})=(\phi \mbox{ mod } 2\pi/3,1-|\delta|)$ for $\delta \geq0$ and $((\phi+\pi) \mbox{ mod } 2\pi/3,1-|\delta|)$ for $\delta < 0$.
For example, the three purple stars in Fig.~\ref{fig:geom} (d) mark points of the parameter space with the same color and, therefore, representing the same bulk lattice (up to a rigid translation). This consideration leads to a peculiar result that is, given a certain bulk lattice configuration (e.g. the red spot in Fig.~\ref{fig:geom} (d)), the lattice can evolve adiabatically into another (e.g. purple stars) by following several different paths in the parametric space that may or may not cross $\delta =0$.
As previously mentioned (and clarified in detail in the following), $\delta =0$ identifies the closing of the bulk band gap. Considering that the closing and reopening of the band gap is an indicator of a possible topological transition, the results in parameter space raises the interesting question about the two configurations belonging or not to the same topological phase.
Along the same line of reasoning, one could wonder if the two lattices belonging to the same bulk pattern and differing, at most, for a relative rigid translation (or rotation) could belong to distinct topologically phases;
given that while crossing the interface between them the band gap closes and reopens. If this latter question admitted a positive answer, then edge states might exist at a dislocation interface between these two phases.
To gain more insight on this topic and answer the above question, the phononic band structure and its topological properties must be analyzed in detail.

\begin{figure}[ht]
\includegraphics[width=.483\textwidth]{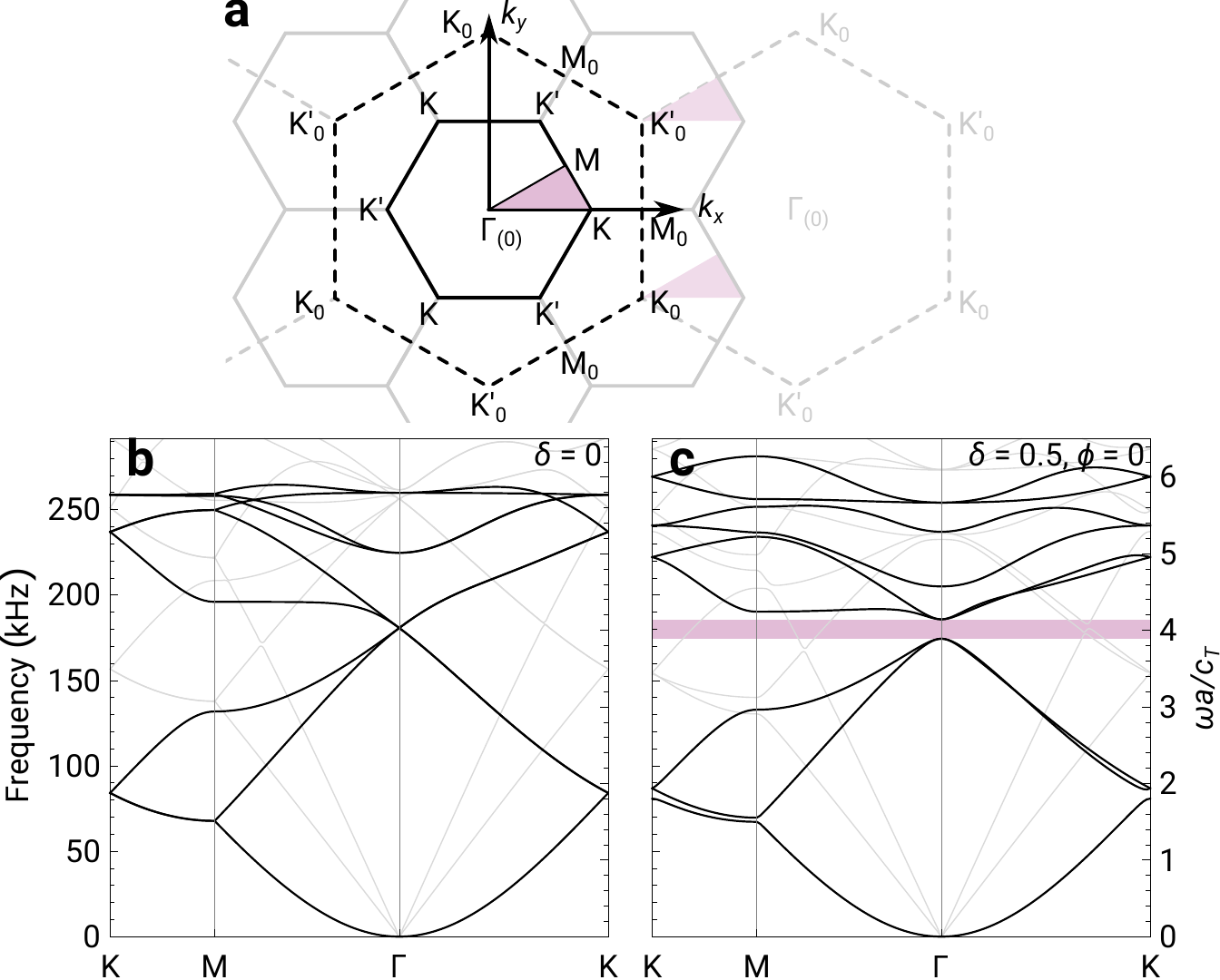}% lattice geometry
\caption{\label{fig:bndstr}
(a) Schematic of the reciprocal space illustrating the formation of the double Dirac cone by zone folding. The solid black hexagon indicates the first Brillouin zone (BZ) associated with the supercell and the black dashed hexagon shows the first BZ related to the original primitive cell. The Dirac cones at the two valleys ${\rm K}_0 (\frac{2\pi}{\sqrt{3}a_0},\frac{-2\pi}{3a_0})$ and ${\rm K}'_0 (\frac{2\pi}{\sqrt{3}a_0},\frac{2\pi}{3a_0})$ of the primitive cell move to and overlap at the $\Gamma$ point of the supercell. (b) The phononic band structures of the reference lattice ($\delta=0$) and (c) of a Kekul\'e distorted lattice with $\delta = 0.5, \phi =0$.
}
\end{figure}

Before introducing the Kekul\'e distortion (that is considering an unpertubed lattice having $(\delta=0, h_1=h_2=h_3=h_0)$), the ``unit cell'' in the inset of Fig.~\ref{fig:geom} (b) is a supercell, and its original primitive lattice is characterized by basis lattice vectors having $1/\sqrt{3}$ length and rotated of $\pi/2$ with respect to those of the supercell. The band structure associated with this supercell exhibits a double Dirac cone at the $\Gamma$ point (Fig.~\ref{fig:bndstr} (b)) that is the result of the folding of the two cones at the valleys ${\rm K}_0 (\frac{2\pi}{\sqrt{3}a_0},\frac{-2\pi}{3a_0})$ and ${\rm K}'_0 (\frac{2\pi}{\sqrt{3}a_0},\frac{2\pi}{3a_0})$ of the original primitive cell. This concept is visually exemplified in Fig.~\ref{fig:bndstr} (a) where the solid black hexagon indicates the first Brillouin zone (BZ) associated with the supercell and the black dashed hexagon shows the first BZ of the original primitive cell. The gray hexagons indicate the $\mathbf{k}$-space duplicates of the BZs. The original Dirac degeneracies at the valleys ${\rm K}_0$ and ${\rm K}'_0$ of the primitive cell were a consequence of intact space inversion symmetry (SIS) and TRS, while the new four-fold degeneracy at the $\Gamma$ point of the supercell is an artifact of band folding and merely due to the artificial selection of the supercell. However, the supercell could be taken as the new reference configuration (i.e. the primitive unit cell) when a local perturbation (such as the Kekul\'e distortion) is applied to it. The symmetry breaking resulting from the local perturbation would lift the four-fold degeneracy opening a band gap at the $\Gamma$ point.

Figs. \ref{fig:bndstr} (b) and (c) show the phononic band structures of the reference lattice ($\delta=0$) and a distorted one ($\delta = 0.5, \phi =0$), respectively. The band structure was obtained by numerically solving the elastodynamic governing equations for the Bloch eigenstates. The black curves in the plots correspond to antisymmetric guided Lamb ($A$) modes, which are essentially flexural modes of the plate waveguide. The light gray curves show the symmetric ($S$) modes. %, and they are combinations of the fundamental symmetric Lamb ($S_0$) modes and the fundamental shear horizontal ($SH_0$) plate modes. 
Given the phononic plate is symmetric with respect to its neutral plane, the flexural modes are completely decoupled from the symmetric modes and from now on we will focus only on the flexural modes.

In Fig.~\ref{fig:bndstr} (b), the double Dirac cone with four-fold degeneracy shows at the $\Gamma$ point, and the frequency bands are doubly degenerate throughout $\Gamma$-K. However, they splits along the $\Gamma$-M direction. This is due to the anisotropy of the band structure near the valleys ${\rm K}_0$ and ${\rm K}'_0$. Indeed, it is known that the dispersion around the valleys is only isotropic under linear $\mathbf{k}$ approximation, and shows trigonal warping away from the valleys \cite{trigonalwarping2000}. The two mirrored trigonally warped cones cross each other.
The frequency spectra remain degenerate along the $\Gamma$-K section, corresponding to the original ${\rm K}_0$-${\rm K}_0$ and ${\rm K}'_0$-${\rm K}'_0$ directions (which happens to be the intersecting part of the two warped cones), while split along the $\Gamma$-M direction that corresponds to the original ${\rm K}_0$-$\Gamma$ and ${\rm K}'_0$-${\rm K}_0$ directions
    (see Sec.~\ref{app:pseudospin} and Fig.~\ref{fig:split} in the Supporting Information (SI) for a drawing helping the visualization of this band structure).
This observation already indicates a major discrepancy between the phononic ``zone-folding'' system and the electronic 2D topological insulators. For the latter, in the absence of inversion asymmetry, and spin-orbital coupling, the entire energy band structure are doubly degenerate. On the contrary, for phononic systems, the two-fold degeneracy only occurs along six discrete directions ($\Gamma\mbox{-K}$ and $\Gamma\mbox{-K}'$).
    Note that, although these lines of intersection connected at the $\Gamma$ point might be reminiscent of exceptional points, the degeneracy at the $\Gamma$ point is still a diabolic point given that the system is Hermitian \cite{ozdemir2019parity}.

In Fig.~\ref{fig:bndstr} (c), under the effect of the Kekul\'e distortion, the four-fold degeneracy is lifted and a band gap opens up, as indicated by the pink shaded box. On the other hand, the spectra along $\Gamma$-K, are now lifted with the only exception of an isolated two-fold degeneracy at the $\Gamma$ point. We anticipate that this behavior could be interpreted as an effective pseudospin-orbit coupling that splits the two pseudospin bands.
We should also note that, in electronic systems, even in presence of spin-orbit coupling, the Kramers theorem requires that double degeneracies occur both at the $\Gamma$ and the three M points (given that the M points are also invariant under $\mathbf{k}\rightarrow-\mathbf{k}$ as they are midpoints between two neighboring $\Gamma$ points) \cite{Z2}. Obviously the ``zone-folding'' system does not follow the same behavior, and shows significant splitting at the M point (Fig.~\ref{fig:bndstr} (b) and (c)). This again points out the discrepancy between the two systems.

    In fact, these substantial differences can be ascribed to different symmetry operators that protect the degeneracy at the $\Gamma$ point. In electronic QSH systems, time-reversal symmetry and $\mathcal{T}^2=-\mathbf{1}$ guarantee the Kramers degeneracy at time-reversal invariant points such as the $\Gamma$ and M points in the $\mathbf{k}$-space.
    On the other hand, in the current elastic lattice, we show that only the following operators property $(C_3\mathcal{T}_F)^2=-1C_3^2$ holds in the vicinity of the $\Gamma$ point, where the operator $\mathcal{T}_F$ takes the expression of the time-reversal operator for spin-$\frac{1}{2}$ fermionic systems.
    The antiunitary operator $C_3\mathcal{T}_F$ maps the system between different azimuth angles in the $\mathbf{k}$-space (i.e., the angle between the $k_x$-axis and the current $\mathbf{k}$-vector), and the $\Gamma$ point becomes the $(C_3\mathcal{T}_F)$-invariant self-dual point \cite{duality}, which leads to the degeneracy at the $\Gamma$ point. See further details on this point in Sec.~\ref{app:kramers} in SI.
    
    This crucial difference ultimately affects how the topological index of a bulk lattice can be defined. Similarly to the TKNN number \cite{TKNN} that cannot classify nontrivial electronic QSH systems from ordinary ones, the typical $\mathbb{Z}_2$ index for electronic QSH system also becomes inappropriate for the current system (since it is based on $\mathcal{T}^2=-\mathbf{1}$). Instead, we define a local topological order that resembles the $\mathbb{Z}_2$ index in that it is based on integrals of the pseudospin-resolved Berry curvature localized near $\Gamma$ (pseudospin-Chern number), as described in Sec.~\ref{app:nu}. The reader might doubt the legitimacy of this approach under the scenario of pseudospin coupling. We show in Sec.~\ref{app:CNM} of SI that the even in the presence of pseudospin coupling, the off-diagonal elements of the Chern number matrix vanish due to odd $\mathbf{k}$-space symmetry of the inter-pseudospin Berry curvature and the result is the same as when the pseudospins decouple.

One of the standard procedures to analyze the system in proximity of the degeneracy is through the $\mathbf{k}\cdot\mathbf{p}$ approach (or similar low-order perturbation approaches), which expands the eigenstates around the $\Gamma$ point by using a basis formed by the degenerate eigenstates. This procedure leads to a description of the system in the form of a $4\times 4$ block diagonal Hamiltonian that is quadratic in $\mathbf{k}$. After the application of a proper change in basis \cite{kekuleprl2017,kekule2019}, this Hamiltonian can be mapped to the Bernevig-Hughes-Zhang (BHZ) model of topological insulators. This simplified Hamiltonian yields four frequency bands that are quadratic in $\mathbf{k}$ with the two upper and the two lower bands being doubly degenerate, respectively. The basis used to obtain this form of the Hamiltonian defines the pseudospin eigenstates.
%They can also be constructed by artificially assembled in-plane circular polarized modes from degenerate eigenmodes with parity symmetry (dipole or quadrupole patterns associated with $p$- and $d$- orbitals \cite{photonicgapped2015}).

Typically, the perturbed lattices are classified as ``nontrivially'' gapped materials if they satisfy either one of the following criteria: 1) the spectra of the $p$- and $d$-orbitals are inverted \cite{photonicgapped2015},
or 2) $B/M<0$ (where $B$ and $M$ are parameters from the BHZ Hamiltonian) \cite{AcousticTIAir}. However, it can be seen (details provided in SI) that in a continuous and periodic medium, the parity of the wavefunctions used in classifying $p$- or $d$-orbitals depends on the reference frame.
It is also reported that the mapping to the BHZ model actually depends on the gauge choice or the unit cell selection \cite{kekuleprl2017,kekule2019}, therefore the possible ``nontrivial'' character of the lattice is indeterminate. On the other hand, unlike nontrivial 2D topological insulators that support gapless helical edge states on the boundary, the phononic lattices classified as ``nontrivial'' still do not support edge states on its boundary with either vacuum or air. In these materials, edge states only exist on the interface between two gapped lattices. Therefore, the classification as a nontrivial material acquires a somewhat more arbitrary character and the conventional bulk-edge correspondence principle does not apply.

In addition, the low-order Hamiltonian cannot capture the band splitting along the $\Gamma$-M direction which is a distinctive feature compared with electronic systems. The band splitting of pseudospin pairs along the $\Gamma$-K direction occurs due to the higher order pseudospin-orbit coupling terms.
These arguments show that the conventional perturbation approach is not sufficient for an accurate description of the phononic Kekul\'e lattice. 
%In the following,

We also carry out a thorough analysis of the topological band properties using \textit{ab initio} calculations (details provided in SI). An explicit expression for the phononic pseudospins
    is provided and numerically evaluated based on the circulation integral of the acoustic Poynting vector throughout the unit cell that tiles periodically covering the entire real $\mathbf{x}$-space. This concept bears analogy with the $\mathbf{k}$-space representation of the acoustic spin and orbital angular momenta discussed in the literature \cite{toftul2019acoustic,bliokh2019spin,burns2020acoustic} (see Sec.~\ref{app:pseudospin} in SI) .

 Results show that the pseudospin indices actually depend on the choice of the reference frame, and consequently the pseudospin-dependent Chern numbers and the local topological order are also non-unique. This observation opens the way to conceive gapless edge states on the interface between lattices having the same bulk pattern but differing up to a relative translation, as illustrated in the following section.

Electronic topological insulators have a pair of \textit{helical} edge states that are topologically protected, counter-propagating, and with gapless energy spectra across the entire band gap.
One of the key element leading to the gapless condition is the Kramers degeneracy guaranteed at $\mathcal{T}$-invariant points in $k_\parallel$-space including $k_\parallel=0$. 
%The gapless condition at the high symmetry points in $\mathbf{k}$-space In 1D BZ for the edge state, these hisymmetry points are the BZ center, $k_\parallel = 0$, and the BZ boundary $k_\parallel = \pm\pi/a$) is guaranteed by Kramers theorem under TRS, $E (\mathbf{k,\uparrow})=E (\mathbf{-k,\downarrow})$. 
On the other hand, the existing literature shows that, in a phononic analog of a topological insulator based on the zone-folding approach, the edge states are generally gapped at $k_\parallel = 0$.
This result is due to the fact that Kramers theorem is valid only for systems with half-integer total spin, such as electronic systems. In phononic systems, the``synthetic Kramers pair'' is created by the pseudospins. When the two pseudospin states are close (in $\mathbf{k}$-space), they tend to repel each other and the eigenstates show strong mixing that completely annihilates the pseudospins. Given that the pseudospin is generally not conserved, the Kramers degeneracy does not hold and the gap at $k_\parallel = 0$ typically appears for the phononic edge states. This observation also means that the edge states are not robust and prone to backscattering since the counter-propagating edge states are already strongly coupled and gapped even before introducing defects.
Previous studies ascribed the presence of the gap to the lack of $C_{6v}$ symmetry of the lattice near the domain wall interface \cite{photonicgapped2015}, and attempted to reduce the gap width by smoothing the transition between the two lattices \cite{zonefold2018}. However, as we find out in the following, the gapless edge state does not strictly require local $C_{6v}$ symmetry and can exist on an abrupt dislocation interface. 

Given the pseudospins have a certain distributed pattern throughout the phononic lattice, the position of the edge which terminates the bulk lattice (or equivalently, the translation of the bulk pattern relative to the edge) is certainly expected to serve as a key parameter and to affect the behavior of the edge state propagating on it.
A proper skipping-orbit condition on the edge can in fact realize nonrestrictive ballistic transport of the phonons and therefore, achieving fully decoupled, counter-propagating, and gapless phononic helical edge states.
In a recent study \cite{kekule2019}, it was shown that this condition can be realized by fine-tuning the parameter $\phi$ of the two adjacent lattices that are space-inverted images of each other.
In the present study, in order to demonstrate that the edge state can exist on a dislocation interface, we let the adjacent lattices have the same bulk lattice pattern so that they can differ only up to a relative translation of amplitude $a_0$ along the $y$-direction. Such translation distance preserves the overall honeycomb pattern across the two lattices which, although not necessary, makes the fabrication of the material sample very easy. Under a fixed reference frame, this lattice assembly can also be considered as integrating two lattices with parameters $\phi$ and $\phi+2\pi/3$. To make it even simpler, we let $\phi=0$, so that the dislocation interface is a plane of mirror symmetry. Once again, these assumptions do not limit the generality of the results but facilitate the final lattice configuration to be fabricated and used for the experiment. Indeed, under these conditions, the edge state will be symmetric with respect to the interface, hence also facilitating its excitation via a single transducer placed right on the interface. 

\begin{figure}[ht]
\includegraphics[width=.483\textwidth]{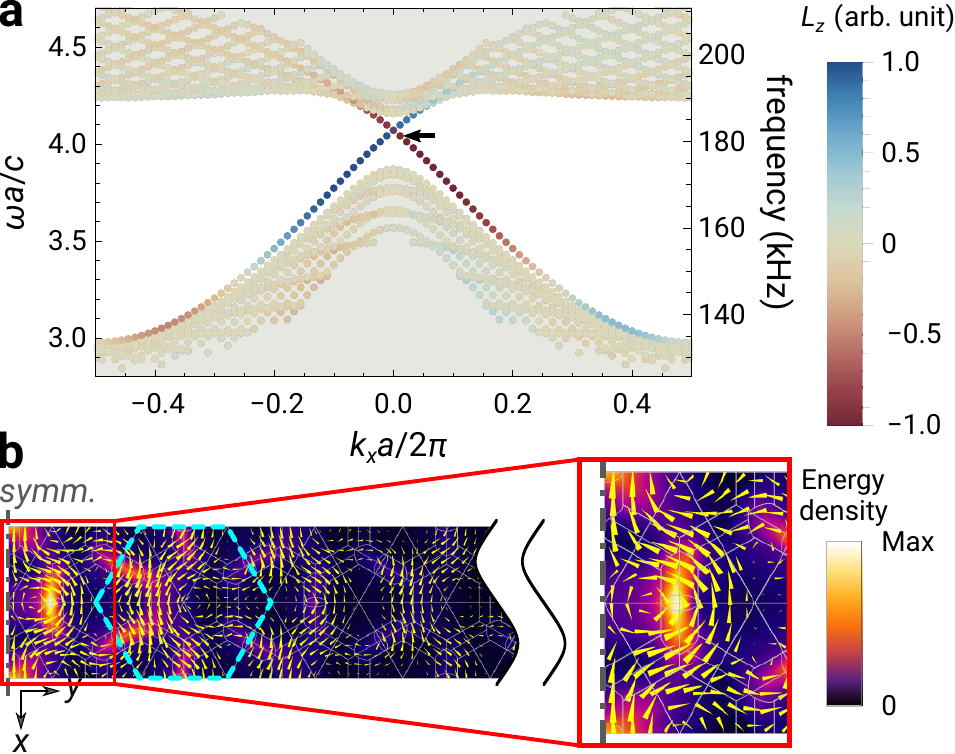}%
\caption{\label{fig:edgedisp} (a) The edge state spectrum along the optimized dislocation interface. The color denotes the pseudoangular momentum. (b) The strain energy distribution (color) and the mechanical energy flux (yellow arrow, log scale) of the eigenstate indicated by the arrow in (a). The dashed cyan hexagon shows the domain used for the pseudoangular momentum integral. The energy flux near the interface indicates a typical skipping-orbit pattern.
}
\end{figure}

%A method is reported that by tuning the potential of the outermost atoms (or engineering the edge features) in a quantum (or acoustic) valley-Hall systems, the edge states can be largely tailored from a flat zero mode to a gapless edge state around the valley \cite{edgetune,myqv}.
By modifying the depth of the groove located on the interface, the edge state dispersion can be readily tuned. In particular, by sweeping through different depths, we find the value such that the edge states reaches accidental degeneracy at $k_\parallel=0$ (see SI for details). Fig.~\ref{fig:edgedisp} (a) shows the edge state spectrum along the optimized dislocation interface, and Fig.~\ref{fig:edgedisp} (b) shows the strain energy distribution and the mechanical energy flux of the eigenstate indicated by the arrow in Fig.~\ref{fig:edgedisp} (a).
The color in Fig.~\ref{fig:edgedisp} (a) denotes the pseudoangular momentum of each eigenstate, whose integral is calculated over a complete hexagonal cell located just next to the dislocation,
as shown in the dashed cyan hexagon in Fig.~\ref{fig:edgedisp} (b).
These results also show that the two edge states are indeed pseudospin polarized, fully decoupled, counter-propagating and gapless, therefore a successful realization of a synthetic Kramers pair. The mechanical energy flux near the symmetric dislocation interface in Fig.~\ref{fig:edgedisp} (b) also indicates a typical skipping-orbit pattern.
Note that the edge state is achieved on an abrupt dislocation, that is without a transition zone smoothing the change in the lattice pattern; in addition,
the unit cells on the dislocation do not possess $C_{6v}$ symmetry.

\begin{figure*}[ht]
\includegraphics[width=1\textwidth]{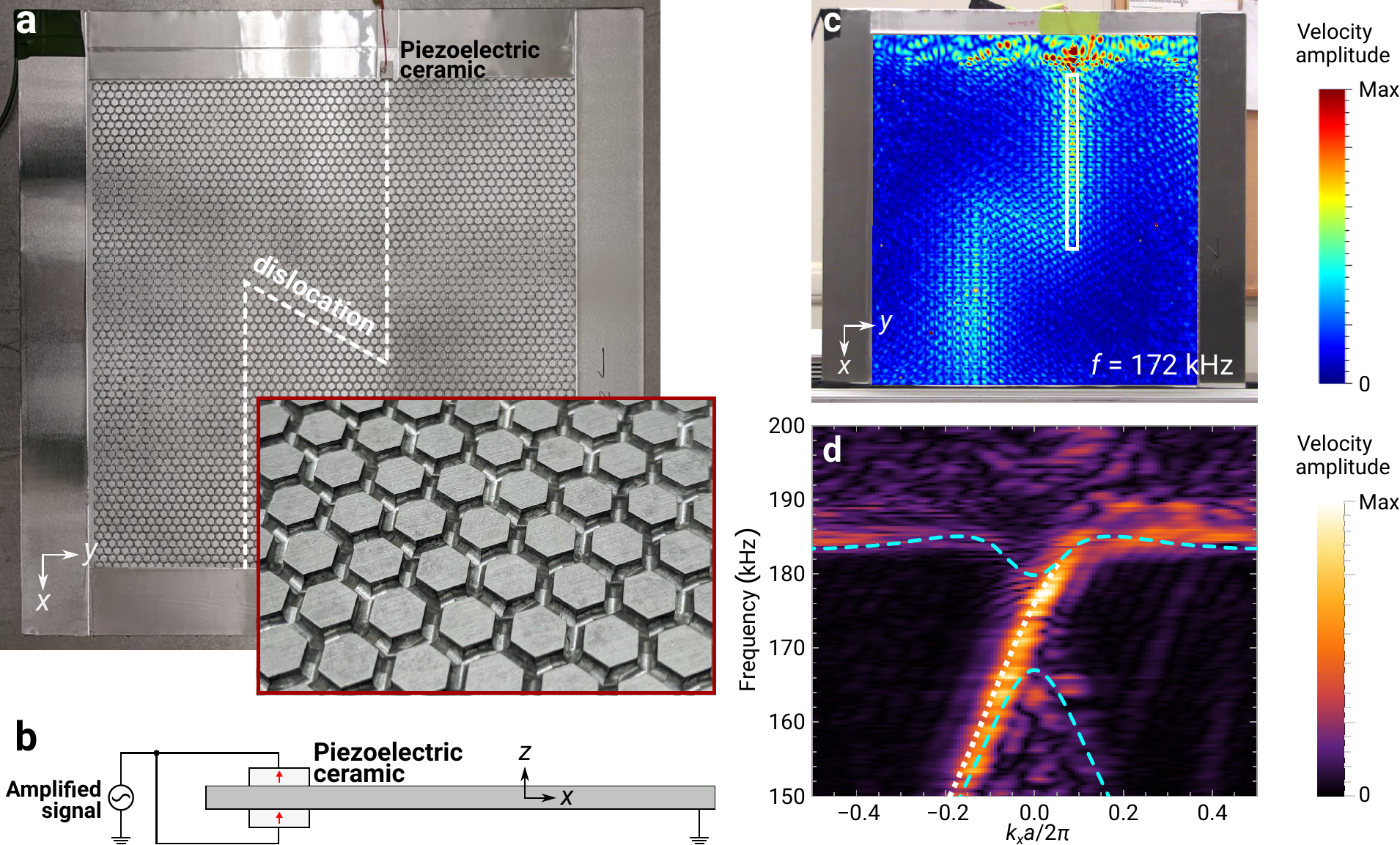}%
\caption{\label{fig:exp}
(a) An image of the fabricated phononic plate. The Z-shape dislocation interface between the two subdomains is indicated by the white dashed line. Piezoelectric ceramic transducers were glued at the top terminal of the Z-dislocation and on both sides of the plate, in order to generate $A_0$ Lamb waves. (b) An illustration of the transducer system and its connections. The red arrows indicate the poling direction of the piezoelectric ceramic plates. %They align in the same $+z$-direction and act out-of-phase as opposite electric field threading them when the voltage signal is applied on their outer electrodes.
(c) The measured velocity amplitude at 172 kHz. The white dashed box includes the data points selected for 1D Fourier transform. 
(d) The $k_x$-spectra in the range 150-200 kHz. The calculated pseudospin-up edge state (cyan dashed curves) and bulk band (white dotted curve) dispersion curves are superimposed to the experimental results. The curves were down shifted by 7 kHz (fractional shift of $-3.8\%$) to match with the experiment data. Results confirm that the edge state along the dislocation is indeed gapless and there is no evidence of coupling to the counter-propagating edge state due to the presence of sharp corners on the path.
}
\end{figure*}

In order to experimentally validate this concept of topological elastic material, we fabricated the phononic plate having two subdomains made of the same bulk honeycomb pattern but differing up to a relative translation. The two domains were used to define a Z-shape dislocation interface (Fig.~\ref{fig:exp}) similarly to the example illustrated in the previous section. Further details on the fabrication and the experimental setup are provided in SI).
Two piezoelectric ceramic patches were glued at the top terminal of the Z-dislocation, on both sides of the plate, as shown in Fig.~\ref{fig:exp} (a, b). The red arrows in Fig.~\ref{fig:exp} (b) indicate the poling direction of the piezoelectric ceramic plates. They were aligned in the same $+z$-direction so to act out-of-phase when applying the voltage on their outer electrodes. This configuration helps an efficient excitation of (primarily) the $A_0$ mode. The response of the plate in terms of the velocity field distribution was measured by a scanning laser Doppler vibrometer.

Figure~\ref{fig:exp} (c) shows the measured velocity amplitude at 172 kHz, which indicates the edge state propagating along the Z-dislocation.
To obtain the edge state dispersion from the measured data, data points along the first segment of the Z-dislocation (enclosed in the white box in Fig.~\ref{fig:exp} (c)) were selected and Fourier transformed so to obtain the $k_\parallel$-spectra (in the range 150-200 kHz) shown in Fig.~\ref{fig:exp} (d). The calculated pseudospin-up edge state (cyan dashed curves) and the bulk band dispersion curves (white dotted curve), previously shown in Fig.~\ref{fig:edgedisp}(a), are superimposed on the spectrum. The curves were downshifted by 7 kHz (fractional shift of $-3.8\%$, normalized by the calculated edge state center frequency) to align with the experimental data. This is a small error considering the operating frequency range and it is likely due to slight deviations of the mechanical properties of the aluminum alloy from the nominal values as well as to small fabrication imperfections. Nevertheless, results confirm that the edge state along the dislocation is indeed gapless across the bulk band gap and that there is no evidence of coupling to the counter-propagating edge state in the presence of sharp corners on the path.

The weak attenuation visible along the consecutive sections of the interface is likely the result of a few factors.
a) There is energy accumulation on the top edge due to the limited absorbing performance of the viscoelastic layer. This accumulation acts as an incoherent source of energy that feeds back into the channel. After this incoherent component of the input is reflected at the first corner, the two remaining branches appear markedly more uniform.
b) The measured response contains not only the edge state, but also the evanescent bulk wave in the band gap. The response of the latter decays exponentially (anisotropically) from the transducers, therefore the mixed response shows larger amplitude on the first section.
c) The decay of the edge state is partially due to the asymmetry of the phononic plate. In the SI, we describe some inconsistencies in the machining results on the two sides which leads to a slight asymmetry with respect to the mid-plane. The asymmetry caused weak coupling between antisymmetric and symmetric modes.
The continuous spectrum of the symmetric modes in the bulk band gap of the antisymmetric modes leads to some energy leakage into the bulk symmetric modes as a result of the weak coupling between $A$ and $S$ modes.
We also conducted a time-transient measurement using a wave packet centered at 172 kHz as the input signal showing that the wave packet propagates through the two acute corners without observable reflections.
A full video of the transient response is also provided in SI \cite{Supp}.

This study presented a fully continuous 2D elastic topological waveguide based on the concept of a Kekul\'e distorted lattice and capable of achieving truly gapless phononic edge states. A combination of theoretical and numerical results showed that this phononic pseudospin Hall system cannot be studied and entirely understood by means of traditional perturbation techniques. Reverting to an approach based on in-depth numerical simulations allowed us to show a discrepancy of the degenerate band structure of ``zone-folding'' systems compared to electronic quantum spin Hall systems. This approach also yielded the fine structure of the spectrum splitting due to pseudospin-orbit coupling caused by inversion asymmetry, also not observable via standard perturbation techniques.
We introduced a definition of the pseudospin that
%augmented definition of pseudospin
allowed all eigenstates in $\mathbf{k}$-space to be labeled clearly and that ultimately revealed a distinctive alternating six-lobe pseudospin texture. By direct observation of the eigenstates, we also isolated the origin of the ambiguity of the pseudospin state, and constructed the $\mathbf{r}$- and $\mathbf{k}$-dependence of the pseudomagnetic field associated with the symmetry breaking.
The ambiguity in the pseudospin state also leads to indeterminate pseudospin dependent Chern numbers. We find that the Kekul\'e lattices have the local topological order $\nu=\pm 1/2$ which suggests the possible existence of edge states.
While phononic gapless edge states at zero momentum had not been observed previously, we showed numerical and experimental evidence of their existence in our elastic phononic system. More specifically, edge states were found to be able to propagate on a dislocation interface of two adjacent identical bulk lattices differing only up to a relative translation. The robustness of these states was examined by propagating elastic wave packets into the dislocation waveguide with no observable backscattering at sharp corners along the path. The experimental results were fully consistent with the dispersion structure predicting gapless edge states and decoupled pseudospin polarizations, therefore suggesting that the two counter-propagating edge states form a successful synthetic Kramers pair.

The approach presented in this study suggests a simple yet robust approach to turn any elastic plate structure into a topological waveguide capable of backscattering protected states while still preserving its structural properties. The possibility of creating such highly controlled states on simple dislocations also greatly simplify both design and fabrication. It is possible to envision the application of this concept to customize the dynamic response of thin wall structures in the context of, as an example, vibration and structure-born noise control, acoustic signal transmission, analog wave filtering.

%\nocite{*}
\vspace{5mm}
\begin{center}
    \textbf{\MakeUppercase{Acknowledgments}}\\
\end{center}
The following work was partially supported by the Office of Naval Research under grant \#N00014-20-1-2608, and by the National Science Foundation (NSF) under grant \#1761423.
\clearpage

\vspace{5mm}
\begin{center}
    \textbf{\MakeUppercase{Supporting Information}}\\
\end{center}
\tableofcontents

\section{Ambiguity in \textit{p}/\textit{d}- orbital classification}
\begin{figure}[ht]
\includegraphics[width=0.483\textwidth]{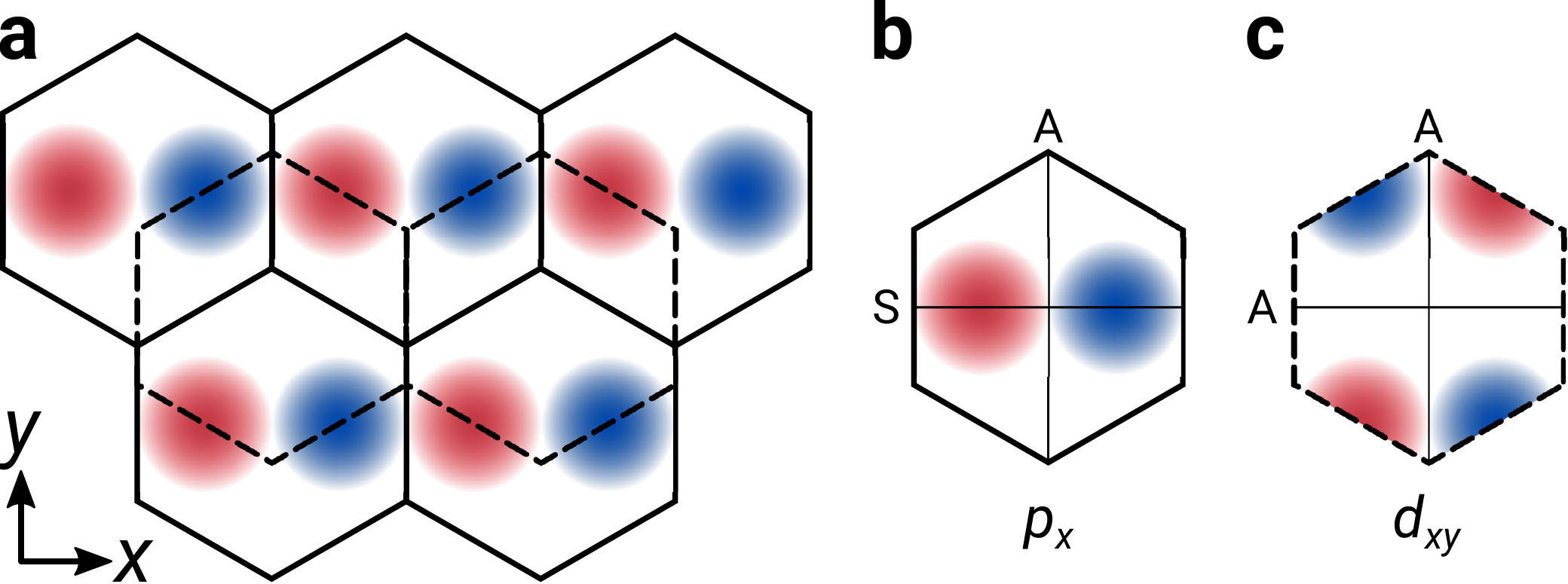}%
\caption{\label{fig:pd}
(a) A periodic wavefunction where red and blue color indicate opposite signs of the wavefunction. The solid and dashed hexagons show two possible choices of the Wigner-Seitz cell. (b) With the solid hexagon chosen as the unit cell, the wavefunction is symmetric with respect to the $x$-axis and antisymmetric with respect to the $y$-axis, therefore the wavefunction has odd parity with respect to the unit cell center, and shows a $p_x$-orbital pattern. (c) When the dashed hexagon is taken as the unit cell, both $x$- and $y$-axes are planes of antisymmetry, and the wavefunction has even parity. The four-lobe alternating pattern of the wavefunction is characterized as a $d_{xy}$-orbital.
}
\end{figure}

In the context of ``zone-folding'' systems, \cite{photonicgapped2015} adopted a criterion to distinguish ``nontrivial'' from ``ordinary'' materials based on the inversion of $p$- and $d$-bands. However, we note that there is an intrinsic ambiguity in the classification of $p$- and $d$-orbitals. Figure \ref{fig:pd} (a) shows an example of a periodic wavefunction where red and blue colors indicate opposite signs of the wavefunction.
\label{app:pdorbitals}

For a continuous periodic lattice such as the considered phononic plate, there is no restriction in choosing the ``atomic site'' as center of the unit cell. In Fig.~\ref{fig:pd} (a), the solid and dashed hexagons show two possible choices of the Wigner-Seitz cell. When taking the solid hexagon, the wavefunction is symmetric with respect to the $x$-axis and antisymmetric with respect to the $y$-axis, therefore the wavefunction has odd parity with respect to the unit cell center, and appears as a $p_x$-orbital-like wavefunction (see Fig.~\ref{fig:pd} (b)). At the same time, by taking the dashed hexagon as the unit cell, both $x$- and $y$-axes are planes of antisymmetry, and the wavefunction has even parity. The four-lobe alternating pattern of the wavefunction is also characterized as the $d_{xy}$-orbital, as shown in Fig.~\ref{fig:pd} (c). Moreover, the general Kekul\'e distorted lattice usually has broken inversion symmetry, so strictly speaking the wavefunctions always have mixed parities.

Such ambiguity does not make this criterion always applicable. Together with the gauge ambiguity in mapping to the BHZ model \cite{kekuleprl2017}, it appears to be meaningless to classify the topological nature of these phononic lattices following the standard approach for electronic topological insulators.

\section{Calculation of the band structure\label{app:band}}
The band structure is obtained by parametrically solving for $\omega$ while varying $\mathbf{k}$ around the irreducible BZ. $\omega$ is obtained as the eigenvalue from the elastodynamic boundary value problem:
\begin{gather}
\label{eq:elasto}
(\lambda+2\mu)\nabla(\nabla\cdot\mathbf{u})-\mu\nabla\times(\nabla\times\mathbf{u})=-\rho\omega^2\mathbf{u},\\
\label{eq:bloch}
\mathbf{u}_{\rm destination}=\mathbf{u}_{\rm source}e^{-i\mathbf{k}\cdot(\mathbf{r}_{\rm destination}-\mathbf{r}_{\rm source})},
\end{gather}
where Eq.~\ref{eq:elasto} is the Navier's equation governing the displacement field $\mathbf{u}(\mathbf{r})$ in the elastic domain, $(\lambda, \mu)$ are the elastic Lam\'e constants, and $\rho$ is the mass density. We assume $(\lambda, \mu)=(50.4, 25.9) {\rm GPa}$ and $\rho=2700 {\rm kg}/{\rm m}^3$ corresponding to an aluminum alloy 6063-T83 which has similar composition to that of the Mic6$^{\mbox{\textregistered}}$ ~aluminum cast plate used in the experiment. Eq.~\ref{eq:bloch} is the boundary condition applied on each of the three pairs of opposite vertical boundaries shown in Fig.~\ref{fig:geom} (b) in the main text to solve for the Bloch eigenstates. The equations are solved using the commercial finite element package COMSOL Multiphysics.

\begin{figure}[ht]
\includegraphics[width=.483\textwidth]{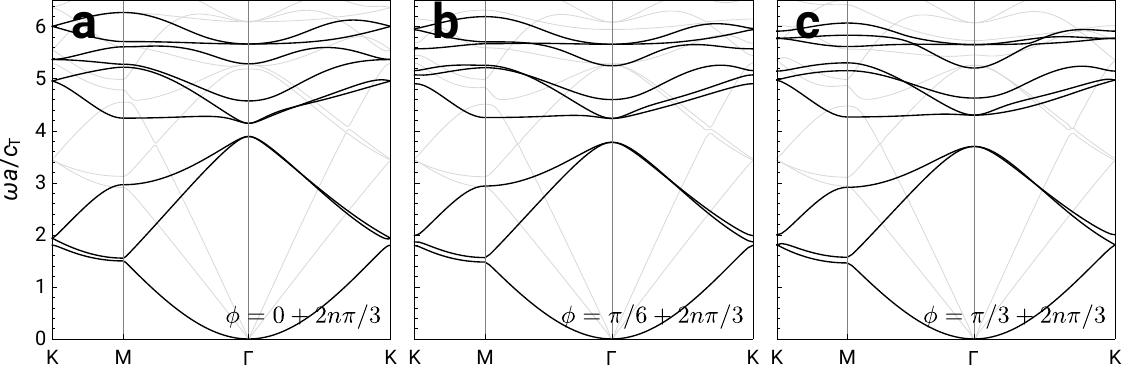}
\caption{\label{fig:BSs} The band structures for lattices having (a) $\phi=0$, (b) $\phi=\pi/6$, and (c) $\phi=\pi/3$, respectively. In all three cases $\delta=0.5$.
%In (a,c) the inversion symmetry is intact and the lattice still possesses $C_{6v}$ symmetry so there are deterministic degeneracies at the K point.}
}
\end{figure}

Fig.~\ref{fig:BSs} shows the band structures for lattices having $\phi=0,\pi/6,\pi/3$, respectively ($\delta\equiv 0.5$).
In the special cases $\phi = 0~{\rm or}~\pi/3 +2n\pi/3, \forall n \in \mathbb{N}$ (Fig.~\ref{fig:BSs} (a,c)), the inversion symmetry is intact, the lattice still possesses $C_{6v}$ symmetry (though the number of $C_6$ axes is reduced to $1/3$ compared with the reference lattice and the original $a_0$-periodicity is also expanded to $a=\sqrt{3}a_0$), and deterministic degeneracies show at the K point. Fig.~\ref{fig:BSs} (b) with $\phi=\pi/6$ represents a more general case in which the lattice symmetry is lowered to $C_{3v}$ and the inversion symmetry is broken. As a result, the degeneracies at the K point are lifted.
We note that, not only the two upper (lower) bands forming the double cone are degenerate at the $\Gamma$ point, but there are also degeneracies involving other bands in the same frequency range that are inseparable from the currently studied double cone system. For example, in Fig.~\ref{fig:BSs} (b,c) along the $\Gamma$-M section, a third frequency band crosses one of the frequency band of the upper cone.
%there are other bands that are inseparable from them.
\begin{figure}[ht]
\includegraphics[width=.35\textwidth]{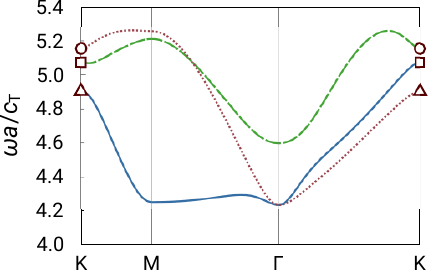}
\caption{\label{fig:BS30up} A close up view of the upper cone of the $\phi=\pi/6$ band structure. The seemingly independent spectra are actually smoothly connected at the K point.
}
\end{figure}
Fig.~\ref{fig:BS30up} shows a zoom-in view of the upper cone of the $\phi=\pi/6$ band structure. In addition to the two bands forming the upper cone (the blue solid and the red dotted curves), there is another band around the same frequency range (the green dashed curve). Due to the broken inversion symmetry, there is no degeneracy at the K point; yet, these seemingly independent bands are smoothly connected at the K point (one can connect the left to the right in Fig.~\ref{fig:BS30up}, hence forming a loop) forming one continuous closed curve. In other words, although the band structure shows three curves that are apparently independent, all states on these three bands can evolve continuously from one to another. A rigorous analysis approach would involve a non-Abelian Berry phase that accounts for interactions between these three bands.
Also, the lifting of the degeneracy induced by the inversion-symmetry-breaking is also expected to contribute additional geometric phase around the K point as was shown in valley Hall systems \cite{fewlayer,bilayerDWexp,bilayerDW,ValleyContrasting,pal2017edge,myqv}. These effects complicate the evaluation of the topological order of the entire band (requiring the integral of the Berry curvature throughout the BZ). We anticipate that, similar to the quantum (acoustic) valley Hall materials, the dynamic behavior (especially the edge state) of the lattice relies on the local geometric phase around the double cone, therefore we can focus the analysis on the neighborhood of the $\Gamma$ point.

\section{Symmetry protected degeneracy and its connection to the Kramers degeneracy} \label{app:kramers}
    Figure~\ref{fig:BSs} clearly illustrated that the use of a general Kekul\'e distortion would result in the formation of two two-fold degeneracies at the $\Gamma$ point that are left behind after lifting the four-fold Dirac degeneracy of the initial unperturbed lattice. In this section we show that the degeneracy can be connected to the existence of a space-time symmetry operator, whose meaning is closely connected to the concept of Kramers degeneracy in electronic systems.

    Recall that, for systems with half-integer total spin, it is always true that $\mathcal{T}^2=-\mathbf{1}$, where $\mathbf{1}$ stands for the identity operator. Whenever the system possesses time-reversal symmetry, the system is at least two-fold degenerate, which is known as Kramers degeneracy \cite{klein1952degeneracy}. The degeneracy is guaranteed by three conditions: i) $\mathcal{T}^2=-\mathbf{1}$, ii) the operator $\mathcal{T}$ is antiunitary \cite{sakurai}, and iii) the time-reversal symmetry of the system is preserved, $\mathcal{THT}^{-1}=\mathcal{H}$ where $\mathcal{H}$ denotes the system Hamiltonian.

    As discussed in the main text, $\mathcal{T}^2$ is always $+\mathbf{1}$ for the present elastic lattice, which means that the Kramers theorem does not apply in a strict sense. However, if an antiunitary operator $\mathcal{A}$ (instead of $\mathcal{T}$) exists and satisfies all the three conditions mentioned above, considerations that are conceptually analogous to the Kramers theorem can be drawn and will still assure the existence of a two-fold degeneracy; this concept was recently demonstrated by Fruchart {\it et al.} \cite{duality}. Specifically, in our system, the operator $\mathcal{A} = C_3 \mathcal{T}_F$, where $C_3$ is an operator corresponding to a proper clockwise rotation in increments of $2\pi/3$ and $\mathcal{T}_F$ is the fermionic time-reversal operator. At the $\Gamma$ point, the operator $\mathcal{A}$ is found to satisfy the same three conditions dictated by the Kramers theorem if the first condition is slightly modified. Indeed, $\mathcal{A}$ satisfies:\\
    \begin{enumerate}
        \item $\mathcal{A}^2 = (C_3 \mathcal{T}_F)^2 = -C_3^2\mathbf{1}$.
        \item $\mathcal{A}$ is antiunitary.
        \item $\mathcal{AHA}^{-1}=\mathcal{H}$.
    \end{enumerate}
    In the following, we show that these three conditions hold and, although representing a modified version of the original Kramers theorem, they still assure the degeneracy at the $\Gamma$ point.

    Assume that in the vicinity of the $\Gamma$ point, the system dynamics of the four bands (the lifted double cones) can be described by a $4\times 4$ Hamiltonian. Although we stated that the low order approximation cannot describe the system dynamics accurately away from the $\Gamma$ point, here we are interested in the degenerate property at the $\Gamma$ point, and the $4\times 4$ Hamiltonian model is again a convenient tool.

    At this stage, we do not know if a degeneracy exists at the $\Gamma$ point for the Kekul\'e distorted lattice. We also do not know the system Hamiltonian around the $\Gamma$ point. Although we can obtain the numerical values of the $4\times 4$ matrix elements by selecting a basis, such as the four degenerate eigenstates of the unperturbed lattice, then expand the perturbed eigenstates and back up the Hamiltonian (see Supplemental Material of \cite{kekule2019}). Instead, simply by examining the symmetry of the eigenstates $\Ket{\mathbf{n}}$ in the neighborhood of the $\Gamma$ point (which they are obtained from the first-principle numerical calculation), we can easily show the three conditions for $\mathcal{A}$ hold.
    %here we need not but only to observe the eigenstates $\Ket{\mathbf{n}}$ in the vicinity of the $\Gamma$ point from the first-principle numerical results for the perturbed lattice, then apply the operator to it, $\mathcal{A}\Ket{\mathbf{n}}$, and we can infer the transformation of the system, $\mathcal{AHA}^{-1}$.

\begin{figure*}[ht]
\includegraphics[width=.8\textwidth]{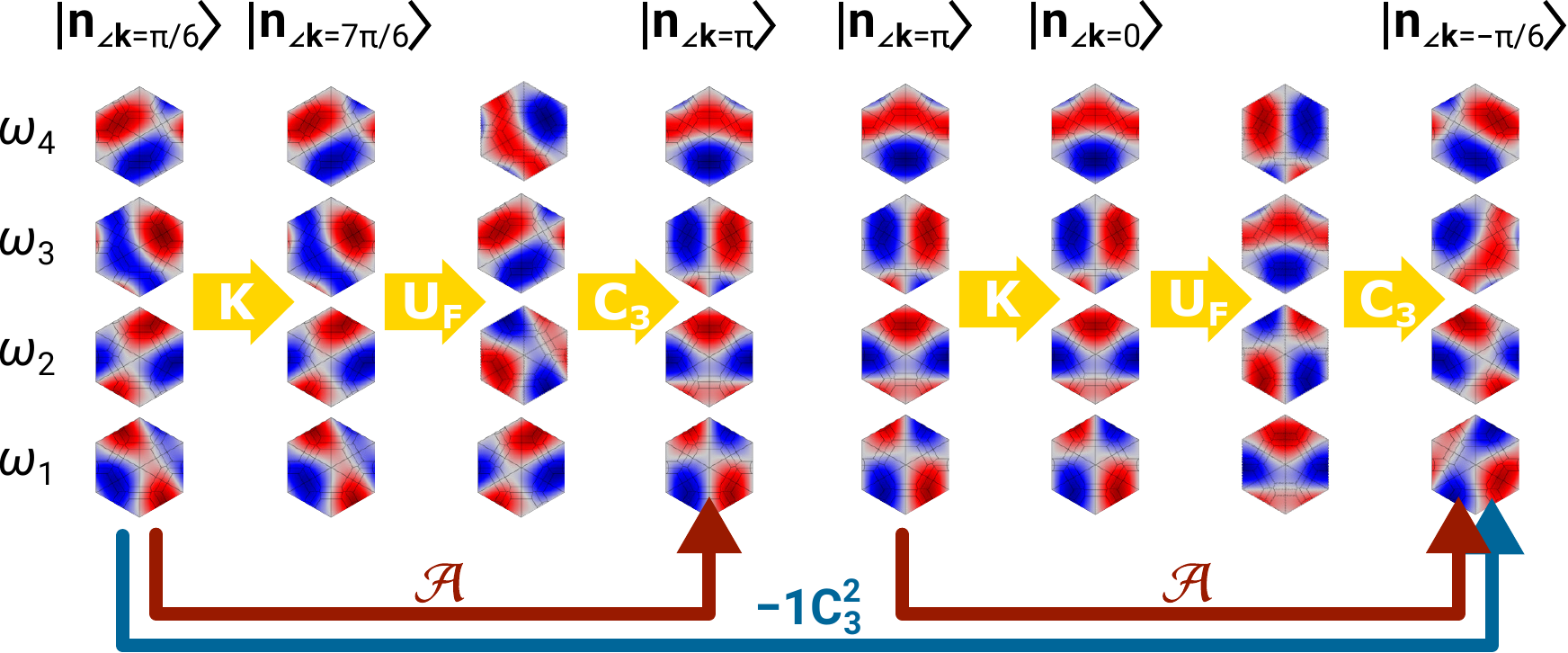}%
\caption{\label{fig:kramers}
        The consecutive application of the antiunitary operator $\mathcal{A}=C_3U_FK$ to the eigenstates at $\|\mathbf{k}\|=10, \angle\mathbf{k}=\pi/6$ (0.035$\Gamma$-M). A single application of the operator maps the original set of eigenstates to those corresponding to the system at $\angle\mathbf{k}=\pi$. A second application of the same operator maps them to the eigenstates corresponding to the system at $\angle\mathbf{k}=-\pi/6$, that is the eigenstates correspond to the original set rotated counterclockwise by $2\pi/3$ and with an additional negative sign. The wavefunctions shown here are the periodic part of the Bloch state of $z$-displacement on the min-plane of the plate, in a Wigner-Seitz cell. 
}
\end{figure*}

    To be more specific, let us examine the four eigenstates in the vicinity of the $\Gamma$ point along the $\pi/6$ azimuth angle in $\mathbf{k}$-space ($\lVert \mathbf{k} \rVert = 10 \mathrm{m}^{-1} = 0.035 \Gamma\mbox{-M}$) as shown in the leftmost column in Fig.~\ref{fig:kramers}.
    The four eigenstates $\Ket{\mathbf{n}}, n=1,2,3,4$ show the ``periodic parts'' of the $n$th Bloch states, on their mid-plane $z$-displacement.

    A lattice with general Kekul\'e distortion possesses $C_{3v}$ symmetry. Taking $\mathbf{k}$ parallel to one of the three planes of mirror symmetry ($\angle\mathbf{k}=\pi/6$), the system Hamiltonian at the specific $\mathbf{k}$ point also inherits that specific mirror symmetry. It follows that the eigenstates are either symmetric or antisymmetric about the symmetry plane, as shown by the numerical results in the first column of Fig.~\ref{fig:kramers}. On the other hand, if the propagation vector $\mathbf{k}$ is perpendicular to a symmetry plane (for example, along $\Gamma\mbox{-K or K}'$, $\angle\mathbf{k}=0,\pi$), the eigenstates also feature
    the same (anti-)symmetry about the mirror plane
    when $\lVert\mathbf{k}\rVert$ approaches 0,  (examples include the well-known acoustic $(1,1)^T$ and optical $(1,-1)^T$ branches of a diatomic phonon lattice as $k\rightarrow 0$ \cite{oxford}). The numerical results shown in the 4th and 6th columns in Fig.\ref{fig:kramers} again highlight the symmetry of the eigenstates.

    We can define explicitly the operator $\mathcal{A}=C_3\mathcal{T}_F$ to be applied to a wavefunction $\Ket{\psi}=c_n\Ket{\mathbf{n}_{\angle\mathbf{k}=\pi/6}}\dot{=}(c_1,c_2,c_3,c_4)^T$, under $\{\Ket{\mathbf{n}_{\angle\mathbf{k}=\pi/6}}\}$-basis representation.
    To get the expression of $\mathcal{T}_F$, we first recall that the time reversal operator is an antiunitary operator $\mathcal{T}=UK$, where $U$ is a unitary operator and $K$ is the complex conjugate operator. For spinless systems, such as our classical elastic system, it is simply $\mathcal{T}=K$. For spin-$\frac{1}{2}$ fermionic systems, $U=e^{i\theta}\sigma_y$ (under $z$-spin representation) where $\theta$ is an arbitrary phase and $\sigma_y$ is the Pauli $y$ matrix \cite{sakurai}. Let $\theta = \pi/2$,
    \begin{equation}
        U=e^{i\pi/2}\sigma_y=
        \begin{pmatrix}
        0 & 1\\
        -1 & 0
        \end{pmatrix}.
    \end{equation}
    By replacing the $z$-spin basis by each (symmetric and antisymmetric) pair of our current $\{\Ket{\mathbf{n}}\}$-basis, we can obtain the analogue {\it fermionic} time reversal operator for our classical mechanical system, $\mathcal{T}_F=U_F K$, where $U_F$ is a block-diagonal matrix,
    \begin{equation}
        U_F=
        \begin{pmatrix}
         0 & 1 & 0 & 0\\
        -1 & 0 & 0 & 0\\
         0 & 0 & 0 & 1\\
         0 & 0 &-1 & 0\\
        \end{pmatrix}.
    \end{equation}
    Note that $\mathcal{T}_F$ is neither the time-reversal nor a symmetry operator of our system. $\mathcal{A}$ is obtained by taking the operator $\mathcal{T}_F$ followed by a $C_3$-rotation. Although the operator $\mathcal{T}_F$ works in the four dimensional $\{\Ket{\mathbf{n}}\}$-eigenspace and $C_3$ in the real physical space (under $\{\Ket{\mathbf{n}}\}$-basis representation, $C_3\dot{=}diag(C_3,C_3,C_3,C_3)$), the operator $\mathcal{A}=C_3 U_F K$ is still antiunitary since $C_3 U_F$ is unitary.

    We can now investigate the effect of the application of the operator $\mathcal{A}$ on a wavefunction $\mathcal{A}\Ket{\psi}=C_3 U_F K\Ket{\psi}$. The process is graphically explained in Fig.~\ref{fig:kramers}. Starting from the four eigenstates in the first column, the application of $K$ maps them to their time-reversed counterpart, i.e., from eigenstates with $\angle\mathbf{k}=\pi/6 \mbox{ to } 7\pi/6$. Next, the operator $U_F$ adds a negative sign to eigenstates $\Ket{1}$ and $\Ket{3}$, and swaps the pairs $\Ket{1}\leftrightarrow\Ket{2}$ and $\Ket{3}\leftrightarrow\Ket{4}$, as visible by comparing the second and third columns of Fig.~\ref{fig:kramers}. Afterwards, the $C_3$ operator rotates clockwise each eigenstates about the lattice's $C_3$ axis by $2\pi/3$, and yields the fourth column in Fig.~\ref{fig:kramers}.
    It can be shown, by direct comparison with the numerical eigenstates, that
    this latter column contains exactly the eigenstates for small $\|\mathbf{k}\|$ along $\angle\mathbf{k}=\pi$. Therefore, $\mathcal{A}$ maps the eigenstates $\Ket{\mathbf{n}_{\angle\mathbf{k}=\pi/6}}$ to $\Ket{\mathbf{n}_{\angle\mathbf{k}=\pi}}$. From now on, for the sake of brevity, ``$\angle\mathbf{k}=$'' will be omitted, so that we can write more compactly,
    \begin{equation}
        \begin{split}
            \mathcal{A}\Ket{\mathbf{n}_{\pi/6}} &= \Ket{\mathbf{n}_{\pi}},\\
            \mathcal{A}^{-1}\Ket{\mathbf{n}_{\pi}} &= \Ket{\mathbf{n}_{\pi/6}}.
        \end{split}
    \end{equation}
    Multiplying $\mathcal{A}$ from the left to both sides of the eigenvalue equation $\mathcal{H}_{\pi/6} \Ket{\mathbf{n}_{\pi/6}} =\omega_n \Ket{\mathbf{n}_{\pi/6}}$ yields
    \begin{equation}
        \mathcal{AH}_{\pi/6} \Ket{\mathbf{n}_{\pi/6}}
            = \omega_n \mathcal{A} \Ket{\mathbf{n}_{\pi/6}}
            = \omega_n \Ket{\mathbf{n}_{\pi}}.
    \end{equation}
    Also,
    \begin{equation}
        \mathcal{AH}_{\pi/6} \Ket{\mathbf{n}_{\pi/6}}
            = \mathcal{AH}_{\pi/6}\mathcal{A}^{-1}\Ket{\mathbf{n}_{\pi}}.
    \end{equation}
 From the right-hand sides of both equations it is seen $\mathcal{AH}_{\pi/6}\mathcal{A}^{-1} = \mathcal{H}_\pi$, given that we could also write $\mathcal{H}_{\pi} \Ket{\mathbf{n}_{\pi}}= \omega_n \Ket{\mathbf{n}_{\pi}}$. We conclude that $\mathcal{A}$ maps the Hamiltonian between different azimuth angles in $\mathbf{k}$-space for small $\|\mathbf{k}\|$, therefore the $\Gamma$ point. Hence, ($\|\mathbf{k}\|=0$) is a self-dual point \cite{duality} where $\mathcal{A}$ maps the Hamiltonian back to itself. In other words, $\mathcal{A}$ is a symmetry operator for the Hamiltonian at the $\Gamma$ point, $\mathcal{AH}_\Gamma\mathcal{A}^{-1}=\mathcal{H}_\Gamma$, or $\mathcal{AH}_\Gamma=\mathcal{H}_\Gamma\mathcal{A}$.

    At the $\Gamma$ point, let $\mathcal{H}_\Gamma \Ket{\mathbf{n}_\Gamma} = \omega_n \Ket{\mathbf{n}_\Gamma}$, then
    $\mathcal{H}_\Gamma\mathcal{A}\Ket{\mathbf{n}_\Gamma}=\mathcal{AH}_\Gamma\Ket{\mathbf{n}_\Gamma} = \omega_n\mathcal{A}\Ket{\mathbf{n}_\Gamma}$. That is, $\Ket{\mathbf{n}_\Gamma}$ and $\mathcal{A}\Ket{\mathbf{n}_\Gamma}$ are both eigenstates of $\mathcal{H}_\Gamma$ with the same eigenfrequency $\omega_n$. 
    Next, we need to show that $\Ket{\mathbf{n}_\Gamma}$ and $\mathcal{A}\Ket{\mathbf{n}_\Gamma}$ are distinct eigenstates, so that there is indeed a degeneracy at $\Gamma$. To show that, we need the condition $\mathcal{A}^2 \Ket{\mathbf{n}_{\pi/6}}=-C_3^2 \Ket{\mathbf{n}_{\pi/6}}$, which can be easily verified by applying $\mathcal{A}$ once more, as demonstrated in Fig.~\ref{fig:kramers}.

    Before we can show that $\Ket{\mathbf{n}_\Gamma}$ and $\mathcal{A}\Ket{\mathbf{n}_\Gamma}$ are distinct, we need to show that for any two-dimensional complex valued function (such as a wavefunction) $\Ket{\psi}$, $-C_3^2\Ket{\psi} \neq \Ket{\psi}$. We can prove it by negation by first assuming $\exists \Ket{\psi}:-C_3^2\Ket{\psi}=\Ket{\psi}$. For brevity, we replace the notation $C_3^2$ ($4\pi/3$ clockwise rotation) by $C_3^{-1}$ ($2\pi/3$ counterclockwise rotation), which are equivalent operations resulting in identical states. It then follows that
    \begin{align}
        C_3^{-1}\Ket{\psi} &= -\Ket{\psi},\\
        C_3^{-2}\Ket{\psi} &= C_3^{-1}(-\Ket{\psi}) = +\Ket{\psi},\\
        C_3^{-3}\Ket{\psi} &= C_3^{-1}(+\Ket{\psi}) = -\Ket{\psi}.\label{eq:contra}
    \end{align}
    The result in Eq.~\ref{eq:contra} is in contradiction to the fact that for all $\Ket{\psi}$, $C_3^{-3}\Ket{\psi}=\mathbf{1}\Ket{\psi}$, so the assumption cannot hold and $\mathcal{A}^2\Ket{\psi} = -C_3^2\Ket{\psi} \neq \Ket{\psi}$ for all $\Ket{\psi}$.

    Finally, we can prove that $\mathcal{A}\Ket{\mathbf{n}_\Gamma}$ and $\Ket{\mathbf{n}_\Gamma}$ are indeed distinct eigenstates. Using once again a negation approach, let's assume that they are the same up to a phase, that is $\mathcal{A}\Ket{\psi} = e^{+i\gamma}\Ket{\psi}$. Due to the antilinearity of the operator $\mathcal{A}$,
    \begin{align}
        \mathcal{A}^2\Ket{\psi}
            &= \mathcal{A}(e^{+i\gamma}\Ket{\psi})\\
            &= e^{-i\gamma}\mathcal{A}\Ket{\psi}\\
            &= e^{-i\gamma}e^{+i\gamma}\Ket{\psi}\\
            &= \Ket{\psi},
    \end{align}
    which leads to a direct contradiction to the previous lemma $\mathcal{A}^2\Ket{\psi} \neq \Ket{\psi}$ for all $\Ket{\psi}$. Hence, $\mathcal{A}\Ket{\mathbf{n}_\Gamma}$ and $\Ket{\mathbf{n}_\Gamma}$ must be distinct eigenstates at the same eigenfrequency, hence they are degenerate. Note that the operator $\mathcal{A}=C_3\mathcal{T}_F$ is a symmetry operator only at the $\Gamma$ point, therefore the degeneracy only show at an isolated point $\Gamma$ rather than indicating globally degenerate eigenstates as considered in previous studies. In summary, the above discussion follows a process analogous to that leading to Kramers degeneracies in spin-$\frac{1}{2}$ systems, but it replaces the time-reversal operator with a space-time symmetry operator (or, equivalenty, a space-symmetry-adapted time operator). It is to highlight this conceptual analogy that the term "synthetic Kramers pair" was used.

The difference between the symmetry operators protecting the two-fold degeneracy in either the current elastic lattice ($(C_3\mathcal{T}_F)^2=-C_3^2$ near $\Gamma$) or the electronic system ($T^2=-\mathbf{1}$ globally) also affects the evaluation of the topological index. This aspect is further discussed in Sec.~\ref{app:nu}.

    A few final remarks are important.
    When two pieces of materials join together, depends on whether the two materials are the same or not, an interface or a dislocation forms between them. In both cases, $C_{3v}$ symmetry is broken (both globally and locally for the cells by the interface/dislocation). As a result, the modified version of Kramers degeneracy is lifted due to the symmetry breaking. This is why in similar photonic and phononic zone-folding systems, the edge states seem to be always gapped  \cite{photonicgapped2015,photonicgapped2018,gapped2018}.
    Nevertheless, with the $C_{3v}$ symmetry broken, we still found a method to recover the degeneracy for the counter-propagating edge states, by tuning the local groove depths $(\delta_{D1},\delta_{D2})$ at the interface (discussed in Sec.~\ref{app:edge}). We note that the attained degeneracy at the interface is not a direct result of the same symmetry operator, since the interface possesses only one mirror symmetry. Rather, we hypothesize that there must exist a symmetry operator, defined in the space of the channel depth parameters $(\delta_{D1},\delta_{D2})$ (see Sec.~\ref{app:edge}), that maps systems described by different geometric parameters; this concept would be analogous to results recently reported by Fruchart {\it et al.} \cite{duality}. For a certain parameter configuration, the system could reach the self-dual state, so that the pseudospin-polarized edge states are decoupled extraordinarily and form degeneracy at zero $k$.

\section{Pseudospins}\label{app:pseudospin}
Unlike electrons having intrinsic spin states, the spin of the phonons can only be well-defined in isotropic media where the longitudinal phonon has spin 0 and the transverse one carries spin 1. In any crystal lattice, provided that the rotation symmetry is no longer a continuous group, purely longitudinal and transverse waves do not exist other than along prescribed directions.
It follows that in a lattice, the phonon spin cannot be defined in a rigorous way, like the electron spin having discrete spin angular momentum \cite{phononspin}.

As mentioned in the previous section, once the Kekul\'e distortion is introduced, the originally degenerate bands along $\Gamma$-K and $\Gamma$-${\rm K}'$ directions are lifted. The resulting inversion asymmetry acts as a pseudomagnetic field and splits the pseudospin-pair bands, $\omega (\mathbf{k,\uparrow})\neq \omega (\mathbf{k,\downarrow})$. Direct observation of the eigenstates obtained from the numerical results confirms that the eigenstates of the two split bands have counter-rotating mechanical energy flux similar to the cyclotron motion of charged particles in a perpendicular magnetic field. 
We define the pseudospins directly using the sign of the pseudocyclotron frequency $\omega_z$ based on the mechanical energy flux circulation. Note that this definition based on direct numerical calculation can be applied to any eigenstate in the $\mathbf{k}$-space, and it is not restricted to doubly degenerate $\Gamma$-K or $\Gamma$-${\rm K}'$ directions. The normalized pseudocyclotron frequency is defined as
\begin{equation}\label{eq:cyc}
    \hat{\omega}_z
    =
    \frac
    {\omega_z}{\omega}
    =
    \frac
    {1}{\omega}
    \frac
    {L_z}
    {I_{zz}},
\end{equation}
where $\omega_z$ is the pseudocyclotron frequency given by $L_z/I_{zz}$, and
$L_z$ and $I_{zz}$ are the $z$-component of the pseudoangular momentum and the pseudomoment of inertia, respectively. These quantities are defined as
\begin{align}\label{eq:Lz}
    L_z &= \int_{\mathrm{cell}}d^3\mathbf{r}~(\mathbf{r}\times \mathbf{J})\cdot \hat{\mathbf{z}},\\
    I_{zz} &=\int_{\mathrm{cell}}d^3\mathbf{r}~(\|\mathbf{r}\|^2-z^2)w.
    \label{eq:Izz}
\end{align}
In the above expressions, $\mathbf{r}=(x,y,z)$ is the position vector with the origin of the coordinates located at the unit cell center,
$\mathbf{J}
%=(J_x,J_y,J_z)
=\frac{1}{2}\mathrm{Re}(\bm{\sigma}\cdot\mathbf{v})$ is the period-averaged mechanical energy flux, while $\bm{\sigma}$ and $\mathbf{v}$ are the stress tensor and the velocity vector, respectively.
In Eq.~\ref{eq:Izz}, $w$ is the period-averaged strain energy density reading $w=\frac{1}{4}\mathrm{Re}(\bm{\sigma}\cdot\bm{\epsilon})$, and $\bm{\epsilon}$ is the strain tensor.
Since the elastic wave is confined in a thin plate, we only consider the energy flux parallel to the $xy$-plane and the associated pseudoangular momentum with respect to the out-of-plane $z$-axis.
Thus, the phonon pseudoangular momentum can only have the $z$-component; this is in contrast with the case of electrons that exhibit an internal spin angular momentum that can potentially point in any direction,  even in 2D materials. We should also note that the definition we employ for the pseudoangular momentum is closer to the concept of orbital angular momentum rather than the internal spin angular momentum in the electronic context.

\begin{figure*}[ht]
\includegraphics[width=0.9\textwidth]{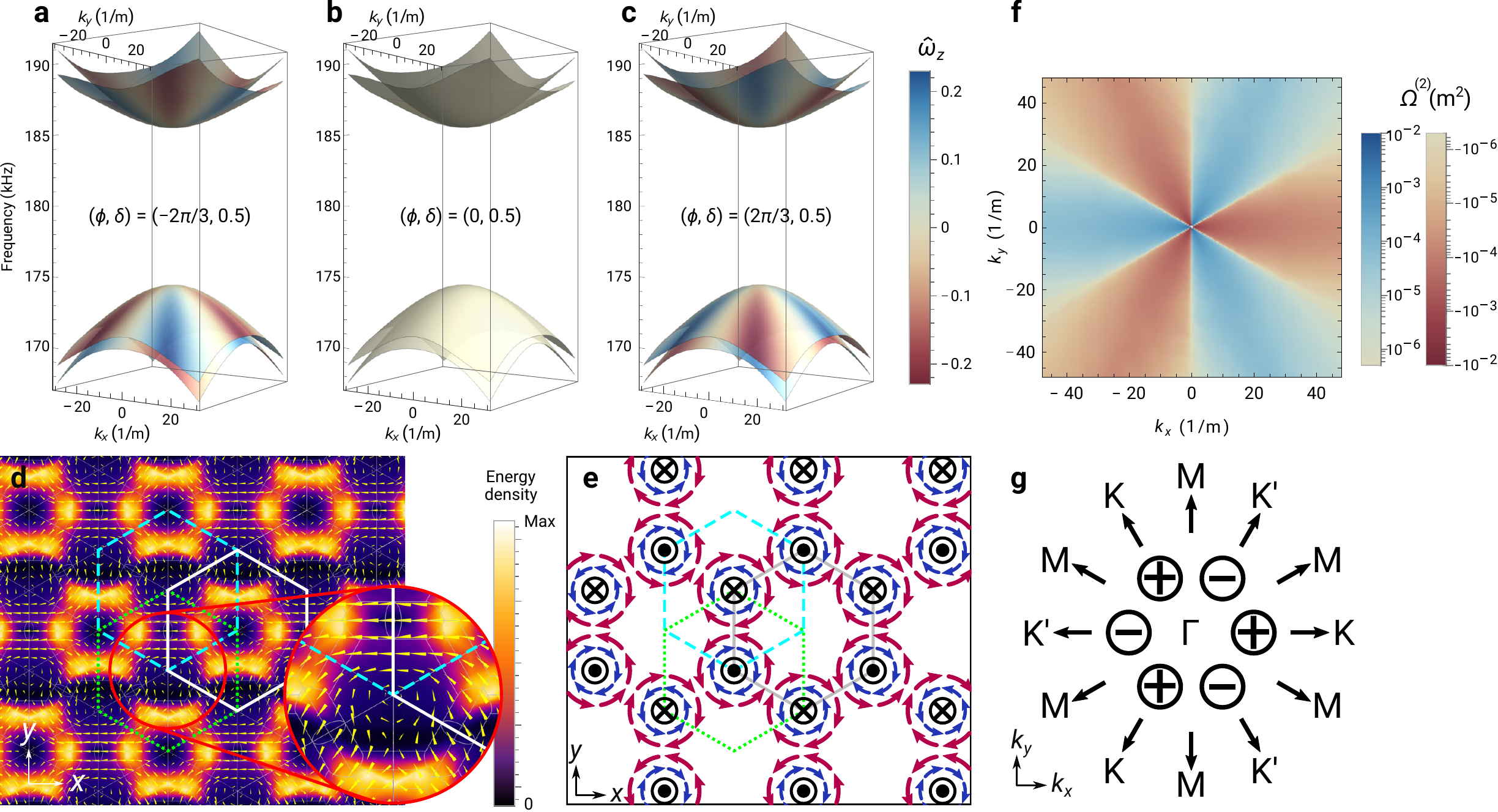}%
\caption{\label{fig:Lz} (a-c) The $\omega$-$\mathbf{k}$ dispersion surfaces near the $\Gamma$ point (0.08 $\Gamma$-K) for $\delta=0.5$ and $\phi = -2\pi/3, 0, 2\pi/3$, respectively. The color marks the pseudocyclotron frequency (with the sign following right-hand rule). (d) Distribution of the strain energy density $w$ along with the mechanical energy flux $\mathbf{J}$ (yellow arrows in logarithmic scale), on the mid-plane of the phononic plate for a state on the second lowest band at $0.026~\Gamma$-K. The three hexagons with cyan dashed, white solid, and green dotted borders mark the Wigner-Seitz cells associated with (a-c), respectively.
(e) Schematic showing the \textbf{r}-dependence of the pseudomagnetic field. The cyan, green, and gray hexagons indicate the three possible choices of the Wigner-Seitz cells centered at a $C_3$ axis. The blue and red circulating arrows illustrate the pseudocyclotron motion of the eigenstate in the first and second lowest bands, respectively, assuming a \textbf{k} point along $\Gamma$-K. The black in- and out-of-plane arrows show the \textbf{r}-dependence of the pseudomagnetic field.
(f) The Berry curvature $\Omega$ of the second lowest band in log scale.
(g) The azimuth-angle-dependence of the pseudomagnetic field in the \textbf{k}-space near the $\Gamma$ point.
}
\end{figure*}

Note that the pseudocyclotron frequency is introduced in order to facilitate the nondimensionalization process and it returns an average spinning frequency, meaning that it accounts for the energy flux of all pseudocyclotrons included in the integral domain.
For simplicity, a Wigner-Seitz primitive cell centered at a $C_3$ symmetry axis (as shown in the inset Fig.~\ref{fig:geom} (b) in the main text) is chosen as the reference domain for integration. The $\omega$-$\mathbf{k}$ dispersion surfaces near the $\Gamma$ point (0.08 $\Gamma$-K) are plotted and color coded using $\hat{\omega}_z$. The results for three configurations identified by the parameters $\delta=0.5$ and $\phi = [-2\pi/3, 0, 2\pi/3]$ are reported in Fig.~\ref{fig:Lz} (a, b, c), respectively.

\begin{figure*}[ht]
\includegraphics[width=0.9\textwidth]{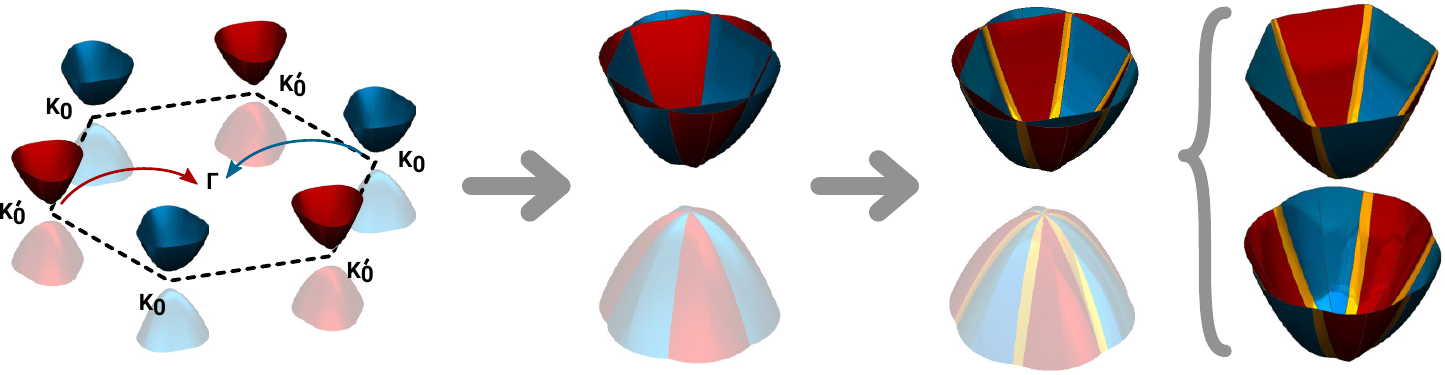}%
\caption{\label{fig:split}
    Conceptual illustration showing the formation of the folded band structure and of the pseudospin coupling in a Kekul\'e distorted lattice.
    First, the double cone composed of opposite pseudospin bands (in red and blue colors) are originated from the two distinct ``original valleys.'' Initially, they are trigonally warped (although the dispersion is circular near the valley, the lattice symmetry only guarantees $C_3$ and $\mathcal{T}$ symmetry, therefore the frequency band has $C_6$ symmetry about $\Gamma$ in the $\mathbf{k}$-space because $\mathcal{T}$ maps $\mathbf{k}$ to $-\mathbf{k}$).
    Second, the folded bands would try to intersect along six different directions but the coupling between the pseudospins repels the neighboring bands preventing any intersection, as indicated by the regions in yellow color. As a result, the two bands are separated except for a degeneracy at the $\Gamma$ point, which is protected by the modified Kramers theorem with the condition $(C_3\mathcal{T}_F)^2=-C_3^2$ as discussed in Sec.~\ref{app:kramers}.
    Ultimately, each of the two bands has a six-lobe alternating pseudospin texture, as found in the numerical results shown in Fig.~\ref{fig:Lz} (a).
}
\end{figure*}

These plots can be leveraged to extract some useful information.
First, we observe that in the presence of Kekul\'e distortion the frequency bands of the upper (lower) two neighboring bands are completely detached from each other except for an isolated degeneracy at the $\Gamma$ point. The repulsion between the two upper (lower) bands, originally subject to trigonal warping, makes each of the four new bands acquire a hexagonal warping.
Second, in Fig.~\ref{fig:Lz} (a, c) it is found that the $\hat{\omega}_z$ pattern for each band shows a six-lobe alternating distribution, in which neighboring frequency bands exhibit opposite $\hat{\omega}_z$ values.
The pseudoangular momentum has extreme values along the $\Gamma\textrm{-K}$ and $\Gamma\textrm{-K}'$ directions, where the spectrum was doubly degenerate before introducing the Kekul\'e distortion, and identically zero along the $\Gamma\textrm{-M}$ directions, where the spectrum was split.
From the pseudocyclotron frequency pattern, we define the pseudospins in the following way: pseudospin up ($\uparrow$) for parts of the frequency band with positive $\hat{\omega}_z$ values, and pseudospin down ($\downarrow$) for those with negative $\hat{\omega}_z$ values. Under such definition, a continuous band is not of a fixed pseudospin but has a six-lobe alternating pseudospin texture. On the other hand, a pseudospin ``band'' can be considered to be composed of parts from two neighboring independently continuous bands. 
    The above discussion can be visualized in Fig.~\ref{fig:split} where, in the frequency bands, red and blue colors indicate opposite pseudospin textures, and the yellow parts illustrate the band repulsion due to pseudospin coupling.

From the above observations, it appears that the BHZ Hamiltonian based on two independent $2\times 2$ massive Dirac Hamiltonians (one per each pseudospin) cannot describe completely our system and will not allow capturing possible geometric phases arising from the interaction between the pseudospin bands. 

Another interesting observation concerns the cases $\phi=-2\pi/3, 0, 2\pi/3$. While, as previously mentioned, they represent the same bulk lattice up to a rigid-body translation, the corresponding pseudoangular momenta are found to be different. Indeed, they have the same magnitude but opposite signs for $\phi=\pm 2\pi/3$, and identically zero values for $\phi=0$. To better illustrate such peculiarity, we plot the strain energy density $w$ along with the mechanical energy flux $\mathbf{J}$ (represented by yellow arrows) on the mid-plane of the phononic plate, as shown in Fig.~\ref{fig:Lz} (d). The plot is taken at $(k_x,k_y)=(10,0)~\mathrm{m}^{-1}$ (along $0.026~\Gamma$-K) for the second lowest phononic band. The plot shows that in addition to an average leftward stream (associated with the negative $x$-gradient of the second lowest band at $0.026~\Gamma$-K), locally circulating patterns distribute repeatedly throughout the lattice. The three hexagons with cyan dashed, white solid, and green dotted borders mark the Wigner-Seitz cells (i.e., the domain of integration) for $\phi=-2\pi/3, 0, 2\pi/3$, respectively.
The inset in Fig.~\ref{fig:Lz} (d) shows a zoom-in near the center of the green hexagon to better visualize the direction of the flux.
Each of the three hexagons covers a full period of the pattern, but either opposite or net zero circulations about the centers are found. Since the reference frames and the location of each pseudocyclotron have
no relative motion (i.e. they are fixed in physical space), the angular momentum of each pseudocyclotron should not vary with different reference frames. One may wonder if these results violate the parallel axis theorem of angular momentum? They do not. The reason is that the pseudocyclotron is periodically distributed, and for each hexagonal domain the integral takes into account portions from different pseudocyclotrons. Then, one may ask why not including all the pseudocyclotron contributions in the same integral domain? It turns out that the improper integral obtained by extending the integral domain in Eq.~\ref{eq:Lz} to infinity does not converge; this is conceptually analogous to the divergent alternating series $\sum_{n=1}^\infty (-1)^n$ that oscillates endlessly. 
The previous discussion shows the intrinsic ambiguity in the pseudoangular momentum so that, for each band with a certain lattice momentum $(k_x,k_y)$, the pseudospin state is only determinate up to a specific choice of the reference frame. 

In summary, in addition to the
discrepancy in the band structure
compared with electronic QSH systems,
the pseudospin indices are dependent on the observer and therefore ambiguous in their definition.
In the result section, we will show that the pseudospin-dependent Chern numbers are in fact gauge dependent. This latter statement might trigger an additional uncertainty and ambiguity in the possibility to define edge states. We will show that it is indeed because of this degree of indeterminacy, due to the dependence on the reference-frame, that we can conceive edge states along a dislocation boundary between two domains of the same bulk lattice.

\section{Berry curvature and pseudomagnetic field}\label{app:B}
In the presence of the two-fold degeneracy at the $\Gamma$ point for opposite pseudospin bands, a non-Abelian Berry phase is considered and the pseudospin-dependent Chern numbers form a $2\times 2$ Chern number matrix (CNM) $C^{\alpha,\beta} = \frac{1}{2\pi}\int d^2 \mathbf{k}~\Omega^{\alpha\beta}$, where $\alpha,\beta = \uparrow,\downarrow$ indicate the pseudospin index \cite{spinChern}. In the above, the overall integral of the inter-band Berry curvature $\Omega^{\alpha\beta}, \alpha\neq \beta$ yields zero, resulting in a diagonal CNM (see Sec.~\ref{app:bc} for details). Therefore, in the following we will only focus on the diagonal elements of the Berry curvature matrix.
Fig.~\ref{fig:Lz} (f) shows the Berry curvature $\Omega^{(2)}$, where the superscript indicates the spectrum index (i.e. the Berry curvature of the second lowest band) in logarithmic scale. The Berry curvature is strongly localized around the $\Gamma$ point where it becomes singular. Interestingly, it also exhibits a six-lobe alternating pattern with extreme values along $\Gamma {\text -}{\rm K}$  and $\Gamma {\text -}{\rm K}'$ and vanishing values along $\Gamma {\text -}{\rm M}$ directions; this pattern is similar to the one observed for the pseudocyclotron motion. 
However, differently from the pseudocyclotrons, the Berry curvature is a geometric property of the frequency band associated with the eigenstate evolution in $\mathbf{k}$-space, which is gauge invariant and does not vary with different reference frames. Such six-lobe pattern matches that obtained for the pseudospins; equivalently, $\Omega^{(2)}$ can be considered composed of $\Omega^{\uparrow\uparrow}$ and $\Omega^{\downarrow\downarrow}$, for each half of the BZ, as the blue and red parts shown in Fig.~\ref{fig:Lz} (f), respectively. Similar considerations can be drawn for $\Omega^{(1)}$ by considering inverted pseudospin patterns.
The Berry curvature is known as the $\mathbf{k}$-space magnetic field \cite{berry1984}. We find that this concept correlates well with the $\mathbf{k}$-space dependence of the pseudomagnetic field in the real \textbf{r}-space that acts on the cyclotrons and splits the pseudospin frequency spectra. Figures \ref{fig:Lz} (e) and (g) show a schematic illustration of both the \textbf{r}- and the \textbf{k}-space dependence of the pseudomagnetic field $\mathbf{B}(\mathbf{k},\mathbf{r})$.

In Fig.~\ref{fig:Lz} (e), the the three cyan, green, and gray hexagons again indicate the three possible choices (identical to those in Fig.~\ref{fig:Lz} (d)) of the Wigner-Seitz cells centered at a $C_3$ axis. The red and blue circulating arrows illustrate the pseudocyclotron motion of the eigenstate in the first and second lowest bands (assuming a \textbf{k} point along $\Gamma$-K), respectively. Given the known pseudocyclotron distribution, we can easily infer the inhomogeneous pseudomagnetic field $\mathbf{B}$, as depicted in black by in-plane and out-of-plane arrows. The pseudomagnetic field must be parallel to the pseudomagnetic moment $\bm{\mu}$ of the pseudocyclotrons of the first band and antiparallel to that of the second band since the shifting of the frequency spectra is $-\bm{\mu}\cdot\mathbf{B}$ (assuming the pseudocyclotron carries positive ``charge,'' $\bm{\mu}$ can be inferred by the right-hand rule). We also note that, for each pseudospin, the total flux through the unit cell is zero, which is analogous to the quantum Hall effect without Landau levels \cite{Haldane}.

Similar to the electron systems, in the phononic pseudospin system the time-reversal partner of a certain eigenstate $\Ket{n(\mathbf{k},\uparrow)}$, is $\Ket{n(\mathbf{-k},\downarrow)}$ (except for the $\Gamma\textrm{-M}$ directions where the spin vanishes). For example, the time-reversal partner of an eigenstate belonging to the second lowest band at some $\mathbf{k}$ point along  $\Gamma {\text -}{\rm K}$, is the eigenstate at the $-\mathbf{k}$ point (along $\Gamma {\text -}{\rm K}'$) with opposite pseudospin. From the pseudospin pattern (Fig.~\ref{fig:Lz} (a)), we know it is still on the second lowest band at the same frequency, which is also a direct result of TRS. It follows that the pseudospins flip, from momentum $\mathbf{k}$ to $-\mathbf{k}$, and also for any $\pi/3$ change in the azimuth angle.
Now, we can back up the $\mathbf{k}$-dependence of the pseudomagnetic field that must switch signs for any $\pi/3$ change in the azimuth angle to yield the opposite spectra splitting for pseudospin-pair states, as shown in Fig.~\ref{fig:Lz} (g).
From the combination of Fig.~\ref{fig:Lz} (e) and (g), we found the $\mathbf{r}$- and $\mathbf{k}$-dependence of the pseudomagnetic field in the current pseudospin Hall system.
In electronic systems, the splitting of two spin bands are usually described by spin-orbit coupling terms. We note that the distinctive six-lobe splitting and pseudospin texture showing here are unprecedented from systems considering conventional Rashba or Dresselhaus spin-orbit coupling terms\cite{meier2007measurement}.

\section{Local topological order}\label{app:nu}
For electronic QSH systems,
the $\mathbb{Z}_2$ topological order can be used to classify the nontrivial topological insulator phase from the ordinary one. The literature provides several different approaches to the calculation of the $\mathbb{Z}_2$ invariant \cite{Z2,spinChern,fu2006time,moore2007topological}. Examples include the use of the wavefunctions of the band pair at the four TR-invariant $\mathbf{k}$ points (i.e., $\Gamma$ and three M points for triangular lattices) that bound half of the BZ \cite{Z2}, the integral of either the Berry connection or the Berry curvature in half of the BZ \cite{fu2006time}, and the difference in the Chern numbers of the opposite spin band in a pair \cite{spinChern,kane2013topological}, $\frac{1}{2}\left(C^{\uparrow\uparrow}-C^{\downarrow\downarrow}\right)~\mbox{mod}~2$ for spin conserved systems. The choice of a specific technique depends on specific symmetries available for the system, as well as on the availability of the Hamiltonian or the band structure.
For the current ``zone-folding'' system, given there is no Kramers degeneracy at the M point as we have seen in Fig.~\ref{fig:bndstr} in the main text (recall from Sec.~\ref{app:kramers} that in our system, $(C_3 \mathcal{T}_F)^2=-C_3^2$ only holds near $\Gamma$), which does not guarantee degeneracy at the M point, we cannot make use of the wavefunctions at those specific $\mathbf{k}$-points to find $\mathbb{Z}_2$.
Also, the idea of making use of half of the Brillouin zone, is originated from $\mathcal{T}^2=-\mathbf{1}$ condition, which might not be suitable for our system.
In fact, we can only focus on the vicinity of the $\Gamma$ point,
to define a \textit{local} topological order that might resemble the $\mathbb{Z}_2$ invariant.

Similarly to elastic valley Hall systems \cite{myqv}, in which the resulting valley dependent Chern numbers are $\pm 1/2$, also in the current system the pseudospin dependent Chern number is half-quantized $C^{\uparrow\uparrow/\downarrow\downarrow}=\pm 1/2 \textrm{ or} \mp 1/2$, where $\pm$ or $\mp$ depends on the choice of the unit cell. This is clearly a direct consequence of the folding which brings the valleys of the initial cell to coincide at the $\Gamma$ point of the distorted cell.
We can therefore define a local topological order, which conceptually follows the same concept of a $\mathbb{Z}_2$ invariant, in the form $\nu=\frac{1}{2}(C^{\uparrow\uparrow}-C^{\downarrow\downarrow})$.

Since the two pseudospin dependent Chern numbers have opposite signs ($\pm 1/2$) and
switch signs simultaneously under different reference frames, the evaluated local topological order
of a specific lattice can be either $\nu=+1/2$ or $-1/2$, depending on the choice of reference frames.
These fractional and gauge dependent values indicate that the defined local topological order is not an exact topological invariant as it is obtained from the local integral of Berry curvature which does not cover the entire effective Brillouin zone \cite{moore2007topological}.
In fact, in the ``zone-folding'' system, we cannot have any topological phase transition in a strict sense as one can easily see from Fig.~\ref{fig:geom} (d) in the main text that a path can be found connecting any two distinct bulk lattices configurations without crossing the abscissa $\delta=0$. In other terms, any two bulk lattices can evolve adiabatically from one to the other without closing the band gap; hence the exact topological invariant does not change.

However, the local topological order is still a useful indicator that we can still follow the similar result of the $\mathbb{Z}_2$ invariants \cite{ReviewKaneTI} to predict the number of Kramers pair edge states on an interface, $N_K=\Delta \nu~\mbox{mod}~2$, where $\Delta \nu$ is the change in the local topological order across the interface.
For example, similar to the electronic topological insulator that a pair of edge states can exist on its outer boundary (i.e., interface between the nontrivial QSH material, $\mathbb{Z}_2=1$, and vacuum, $\mathbb{Z}_2=0$), in the ``zone-folding'' system, a pair of band-gap-crossing edge states can be synthesised on an interface connecting two materials with $\nu_1=+1/2$ and $\nu_2=-1/2$.
At this point, given the sign of local topological order $\nu$ depends on the choice of the reference frame, it is natural to wonder if an edge state can exist on the interface between lattices having the same bulk pattern but with a relative translation (i.e., a dislocation interface).
The answer is indeed affirmative, and a way to conceive a pair of gapless pseudospin-polarized edge states on a dislocation interface is provided in Sec.~\ref{app:edge}.

\section{Pseudospin texture for a general case}
\begin{figure}[ht]
\includegraphics[width=.483\textwidth]{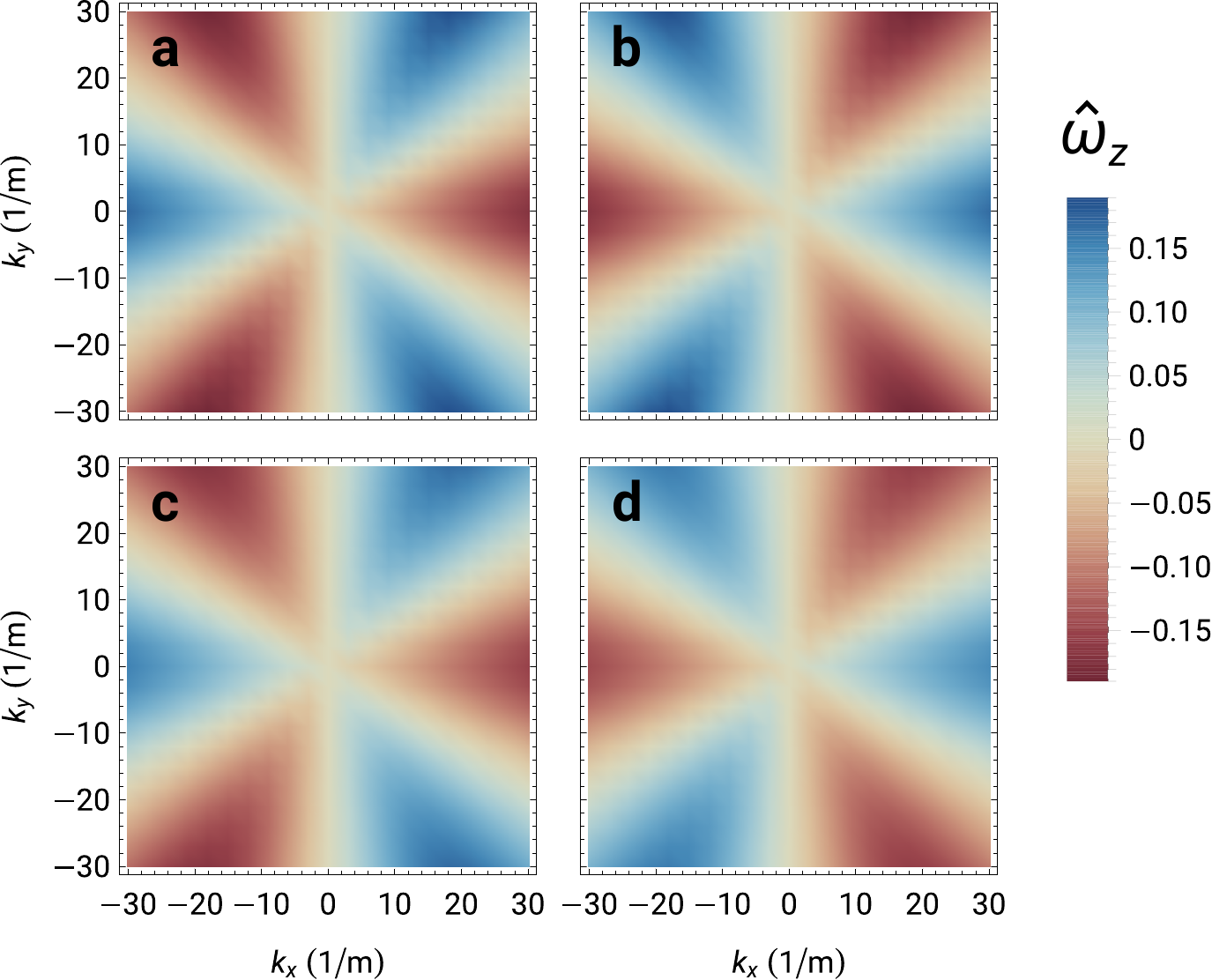}
\caption{\label{fig:Lz30} The pseudospin texture of the lattice with $(\phi,\delta)=(\pi/6,0.5)$. (a-d) shows the normalized cyclotron frequency pattern for bands number 1-4, respectively.
}
\end{figure}

We have shown the alternating six-lobe pseudospin texture for the phononic bands of the lattice with $\phi=0$ ($C_{6v}$). The specific choice of $\phi=0$ was made to create a symmetric dislocation interface and simplify the experimental setup. Fig.~\ref{fig:Lz30} shows the pseudospin texture for the case $\phi=\pi/6$ ($C_{3v}$) that represents a general case with broken inversion symmetry. The pseudospin textures of the four bands still exhibits the same six-lobe pattern, confirming that the symmetric pattern with respect to $k_x$ and $k_y$ has a general character rather than representing the effect of the $C_{6v}$ symmetry of the lattice.
\section{Numerical method for Berry curvature evaluation}
\label{app:bc}
The Bloch eigenstate is obtained from the finite element analysis, in the form of the displacement field $\mathbf{u}^{(n)}(\mathbf{k},\mathbf{r})$, where the superscript denotes the spectrum index.
We denote $\Ket{\mathbf{n}(\mathbf{k})}=e^{-i\mathbf{k}\cdot\mathbf{r}}\mathbf{u}^{(n)}(\mathbf{k},\mathbf{r})$ as the ``periodic part'' of the $n$th Bloch state; similarly $\Ket{\mathbf{m}(\mathbf{k})}$ for the $m$th state.
In the case without degeneracy of the $n$th band, the eigenstate separated by $\delta \mathbf{k}$ from that at momentum $\mathbf{k}$ can be expressed in terms of the eigenstate at momentum $\mathbf{k}$ by accounting for an additional phase, $\Ket{\mathbf{n}(\mathbf{k}+\delta\mathbf{k})} = e^{-i \mathbf{A}^{(n)}\cdot\delta\mathbf{k}} \Ket{\mathbf{n}(\mathbf{k})}$, where $\mathbf{A}^{(n)}$ is the Abelian Berry connection. With the eigenstate being normalized, $\Braket{\mathbf{n}|\mathbf{n}}=1$, we have
\begin{equation}\label{eq:A}
	\Braket{
    	\mathbf{n}(\mathbf{k})|
        \mathbf{n}(\mathbf{k}+\delta\mathbf{k})
        }
=
\Braket{
    	\mathbf{n}(\mathbf{k})|
        \mathbf{n}(\mathbf{k})
        }
        e^{i\mathbf{A}\cdot\delta\mathbf{k}}
=e^{i\mathbf{A}\cdot\delta\mathbf{k}}.
\end{equation}
For infinitesimal $\delta \mathbf{k}$, both sides of the above equation can be simplified (by taking a series expansion up to the linear term in $\delta \mathbf{k}$) as
   \begin{equation}
     \begin{array}{cccccc}
lhs=& 
\Braket{
    	\mathbf{n}(\mathbf{k})|
        \mathbf{n}(\mathbf{k})
        }
        &+ &\Braket{
        \mathbf{n}(\mathbf{k})|
        \nabla_\mathbf{k}|\mathbf{n}(\mathbf{k})}\cdot \delta\mathbf{k}
        &+&O(\delta k^2),
        \\
rhs=& 1 &+ &i\mathbf{A}\cdot\delta\mathbf{k}
        &+&O(\delta k^2).
     \end{array}
   \end{equation}
Comparing the linear terms in both equations, the Berry connection can be written in the well-known form $\mathbf{A}^{(n)}(\mathbf{k})=-i\Braket{\mathbf{n}(\mathbf{k})|\nabla_\mathbf{k}|\mathbf{n}(\mathbf{k})}$. In our numerical implementation, where we must use a small but finite-sized $\delta \mathbf{k}$, we use the form shown in Eq.~\ref{eq:A} that yields better accuracy and numerical robustness when finite difference approximation is used. The finite (central) difference formula to compute the $k_x$ and $k_y$ components of the Berry connection,
\begin{align}\label{eq:AxAy}
\begin{split}
A^{(n)}_x(\mathbf{k})&=
	\frac{\angle \Braket{\mathbf{n}(\mathbf{k}-\delta k\, \hat{\mathbf{k}}_x)|
    	\mathbf{n}(\mathbf{k}+\delta k\, \hat{\mathbf{k}}_x)}}
    {2\delta k},\\
A^{(n)}_y(\mathbf{k})&=
	\frac{\angle \Braket{\mathbf{n}(\mathbf{k}-\delta k\, \hat{\mathbf{k}}_y)|
    	\mathbf{n}(\mathbf{k}+\delta k\, \hat{\mathbf{k}}_y)}}
    {2\delta k},
\end{split}
\end{align}
where ``$\angle$'' extracts the argument of a complex number (mod $2\pi$), $\delta k$ is the step size, and $\hat{\mathbf{k}}_x,\hat{\mathbf{k}}_y$ are the unit vectors in the $\mathbf{k}$-space. The bra-ket inner product sums up the integrals of the three components of the displacement field of the corresponding eigenstates over a unit cell in $\mathbf{r}$-space. 

We note that the eigenstates all have the $U(1)$ gauge ambiguity, and the evaluation of Eq.~\ref{eq:AxAy} requires a smooth gauge choice. To guarantee the smoothness, we pre-process the eigenstates returned from COMSOL Multiphysics with an additional $U(1)$ gauge transformation, $\Ket{\mathbf{n}(\mathbf{k})}\rightarrow e^{i\theta^{(n)}(\mathbf{k})}\Ket{\mathbf{n}(\mathbf{k})}$, where $\theta^{(n)}(\mathbf{k})=\angle u_z^{(n)}(\mathbf{r}_{\rm ref},\mathbf{k})$ is a reference phase which is taken from the argument of the $z$-component of the displacement at a fixed reference point in the unit cell.
%Therefore the phase at every point in the unit cell is taken with respect the reference point.
Given that, for every smooth gauge choice, $u_z^{(n)}(\mathbf{r}_{\rm ref},\mathbf{k})$ must be smooth throughout the $\mathbf{k}$-space, fixing its argument to zero determines a smooth gauge choice. Note that any reference point could be selected, as long as $u_z^{(n)}(\mathbf{r}_{\rm ref},\mathbf{k})$ does not vanish at some $\mathbf{k}$ (hence rendering the Berry connection gauge dependent). The gauge invariant Berry curvature is then obtained as the curl of the Berry connection, also using the finite difference formula,

\begin{align}
\begin{split}
\Omega^{(n)}&=\frac{A_y^{(n)}(\mathbf{k}+\delta k\, \hat{\mathbf{k}}_x)-A_y^{(n)}(\mathbf{k}-\delta k\, \hat{\mathbf{k}}_x)}{2\delta k} \\
&-\frac{A_x^{(n)}(\mathbf{k}+\delta k\, \hat{\mathbf{k}}_y)-A_x^{(n)}(\mathbf{k}-\delta k\, \hat{\mathbf{k}}_y)}{2\delta k}
\end{split}
\end{align}
Considered the degeneracy at the $\Gamma$ point, $\omega^{(1)}=\omega^{(2)}, \omega^{(3)}=\omega^{(4)}$, the non-Abelian Berry connection should be considered. The latter definition takes into account the interaction between two neighboring bands near the $\Gamma$ point and can be obtained by naturally extending Eq.~\ref{eq:AxAy} to form a $2\times 2$ matrix \cite{spinChern},
\begin{align}\label{eq:nonab}
\begin{split}
A^{(m,n)}_x(\mathbf{k})&=
	\frac{\angle \Braket{\mathbf{m}(\mathbf{k}-\delta k\, \hat{\mathbf{k}}_x)|
    	\mathbf{n}(\mathbf{k}+\delta k\, \hat{\mathbf{k}}_x)}}
    {2\delta k},\\
A^{(m,n)}_y(\mathbf{k})&=
	\frac{\angle \Braket{\mathbf{m}(\mathbf{k}-\delta k\, \hat{\mathbf{k}}_y)|
    	\mathbf{n}(\mathbf{k}+\delta k\, \hat{\mathbf{k}}_y)}}
    {2\delta k},
\end{split}
\end{align}
The Berry curvature matrix can be obtained accordingly.

\begin{figure}[ht]
\includegraphics[width=.4\textwidth]{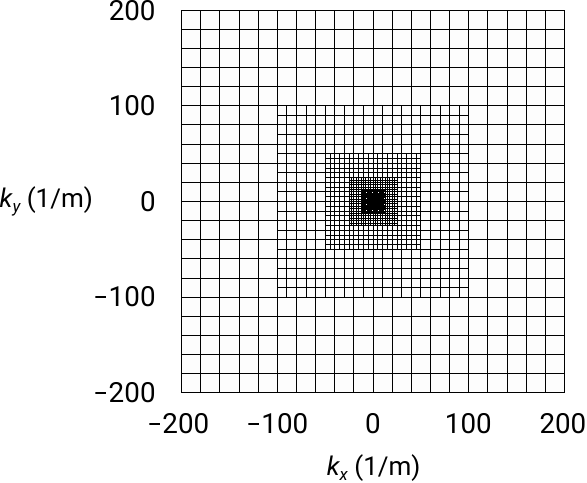}% frame
\caption{\label{fig:frame} The mesh grids used to evaluate the Berry curvature in $\mathbf{k}$-space. Eigenstates, Berry connection and curvature are evaluated at the grid points.
}
\end{figure}

To better resolve the Berry curvature close to the singular $\Gamma$ point, we evaluate the Berry curvature in a step-wise fashion using different step sizes: as getting closer to the $\Gamma$ point, we evaluate Berry curvature using smaller step size in $\mathbf{k}$-space, which it refines by a factor of two for every next level.
%We use an adaptive grid to discretize the $\mathbf{k}$-space. The step size is reduced in a step-wise fashion by a factor of two while approaching the singular $\Gamma$ point.
The refinement is applied for eight levels (enough to guarantee the convergence of the numerical integral) as shown in Fig.~\ref{fig:frame}.

\section{Chern number matrix}\label{app:CNM}
The element of the $2\times 2$ Chern number matrix (CNM) is $C^{\alpha,\beta}=\frac{1}{2\pi}\int d^2\mathbf{k}\ \Omega^{\alpha,\beta} $, where $\alpha$ and $\beta$ denote the pseudospin rather than the spectrum indices. For example, considering the lower cone (spectrum indices $=1, 2$), along $+\Gamma K$, $\Omega^{\uparrow\downarrow}=\Omega^{(1,2)}$. Note that, in an equivalent fashion, we can write $\Omega^{\uparrow\downarrow}=\Omega^{(2,1)}$ along $-\Gamma K$.

The total spin-related Chern number can be explicitly written as $C_t=C^{\uparrow\uparrow}+C^{\uparrow\downarrow}-C^{\downarrow\uparrow}-C^{\downarrow\downarrow}$. We claim, and will show later, that although in our system the pseudospins are not conserved, the off-diagonal elements of the CNM (i.e., $C^{\uparrow\downarrow}, C^{\downarrow\uparrow}$) vanish, therefore we can take the local topological order $\nu=\frac{1}{2}\left(C^{\uparrow\uparrow}-C^{\downarrow\downarrow}\right)$.

Let us consider an off-diagonal element of the Berry connection,
$\mathbf{A}^{\uparrow\downarrow}(\mathbf{k})=-i\Braket{\uparrow(\mathbf{k})|\nabla_\mathbf{k}|\downarrow(\mathbf{k})}$. At some $\mathbf{k}$ point, it is also
$\mathbf{A}^{\uparrow\downarrow}(\mathbf{k})=-i\Braket{\mathbf{m}(\mathbf{k})|\nabla_\mathbf{k}|\mathbf{n}(\mathbf{k})}$. At $-\mathbf{k}$, since the pseudospin indices swapped due to the odd texture, $\mathbf{A}^{\uparrow\downarrow}(-\mathbf{k})=-i\Braket{\mathbf{n}(\mathbf{-k})|\nabla_\mathbf{k}|\mathbf{m}(-\mathbf{k})}$. Observing that $\Ket{\mathbf{n}(-\mathbf{k})}$ is the time-reversed version of $\Ket{\mathbf{n}(\mathbf{k})}$ and we can take
$\Ket{\mathbf{n}(-\mathbf{k})} = \mathcal{T}\Ket{\mathbf{n}(\mathbf{k})} = \Ket{\mathbf{n}^\ast(\mathbf{k})}$ where $^\ast$ means the complex conjugation of a wavefunction or of a scalar, we have
\begin{align*}
\mathbf{A}^{\uparrow\downarrow}(-\mathbf{k})&=-i\Braket{\mathbf{n}(-\mathbf{k})|\nabla_\mathbf{k}|\mathbf{m}(-\mathbf{k})}\\
&=-i\Braket{\mathbf{n}^\ast (\mathbf{k})|\nabla_\mathbf{k}|\mathbf{m}^\ast(\mathbf{k})}\\
&=-i\Braket{\mathbf{n}(\mathbf{k})|\nabla_\mathbf{k}|\mathbf{m}(\mathbf{k})}^\ast\\
&=-i\Braket{\mathbf{m}(\mathbf{k})|\nabla_\mathbf{k}|\mathbf{n}(\mathbf{k})}=\mathbf{A}^{\uparrow\downarrow}(\mathbf{k})
\end{align*}
The above result means that the off-diagonal elements of the Berry connection are even functions, therefore their derivatives ($\partial_{k_p}A_q^{\uparrow\downarrow}, \partial_{k_p}A_q^{\downarrow\uparrow} $ where $p,q=x ~{\rm or}~ y$) must be odd functions. It follows that the off-diagonal elements of the Berry curvature are odd functions, therefore their integrals around the $\Gamma$ point vanish and the CNM in our system is a $2\times 2$ diagonal matrix.

\section{Topology of the \texorpdfstring{$(\phi,\delta)$}{ } parametric space }\label{app:parametric}
\begin{figure}[ht]
\includegraphics[width=.45\textwidth]{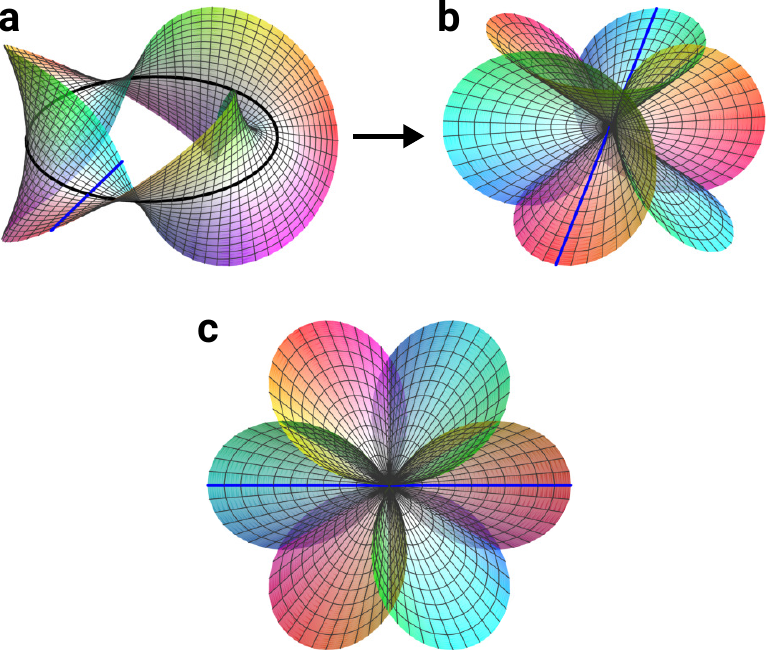}% lattice geometry
\caption{\label{fig:mobius3} A different representations of the $(\delta ,\phi)$ parametric space. (a) The parametric space can be pictured as a 1.5-turn ($3\pi$)-twisted M\"obius strip. The upper and lower bounds of the parameter $\phi \in [0,\pi)$ can be connected to form a loop while at the same time the parameter $\delta \in \left[ -1,1\right]$ is inverted three times, given that the same bulk lattice configuration appears every increased $\pi/3$ in with opposite $\delta$. The blue line indicates $\phi =0$ and the black mid-strip ring is the $\delta =0$ curve. (b) The strip with the mid-ring shrinking to a point, since all points toward the $\delta =0$ curve converge to the same reference configuration. (c) A ``top view'' of (b).
}
\end{figure}

We note that the $(\phi,\delta)$ parametric space shown in Fig.~\ref{fig:geom} (d) in the main text can be visualized in a more natural way (already showing evidence of the six-lobe band properties). The upper and lower bounds of the parameter $\phi \in [0,\pi)$ can be connected to form a loop while at the same time the parameter $\delta \in \left[ -1,1\right]$ is inverted three times, given that the same bulk lattice configuration appears every $\pi/3$ increments with opposite $\delta$. Therefore, the parametric space can be pictured as a 1.5-turn ($3\pi$)-twisted M\"obius strip as shown in Fig.~\ref{fig:mobius3} (a), where the blue line indicates $\phi =0$ and the black mid-strip ring is the $\delta =0$ curve.
Given that all points toward the $\delta =0$ curve converge to the same configuration (that is the undistorted reference lattice), the mid-ring should shrink to a point, as shown in Fig.~\ref{fig:mobius3} (b).  A ``top view'' with the azimuth angle being twice the parameter $\phi$ is shown in Fig.~\ref{fig:mobius3} (c).
Note that an increment of an arbitrary $\Delta \phi$ in $\phi$ does not mean a $\Delta \phi$-rotated lattice or $\mathbf{k}$ vector. As mentioned in an earlier paragraph, for any given $\phi$, the new parameter $\phi + 2\pi /3$ yields a $2\pi /3$ rotation in the entire lattice, or $-2\pi /3$ rotation in the $\mathbf{k}$ vector.
For example, the center of the three red lobes in Fig.~\ref{fig:mobius3} correspond to $\phi =0, 2\pi /3, 4\pi /3$. Therefore, they can also be interpreted as the $\phi =0$ configuration with the reference frame ($\mathbf{k}$ vector) rotated by $0, -2\pi /3, -4\pi /3$. They should all yield the same gauge-invariant Berry curvature. On the other hand, the opposite blue lobes correspond to the same value of $\phi$ with an additional negative sign in $\delta$ which indicates a Berry curvature with opposite sign.
At the same time, the additional negative sign in $\delta$ can also be identically represented by the $\phi =0$ configuration with $\pi$-rotated $\mathbf{k}$ vector.
Therefore, we have linked the distortion parameter $\phi$ with the azimuth angle in the $\mathbf{k}$-space, and the topology of the parametric space already suggests the $\mathbf{k}$-space dependence of the Berry curvature.
%========== above section can be move to supplemental information ==========
\section{Dislocation interface to achieve decoupled gapless edge states}
\label{app:edge}
\begin{figure}[ht]
\includegraphics[width=.483\textwidth]{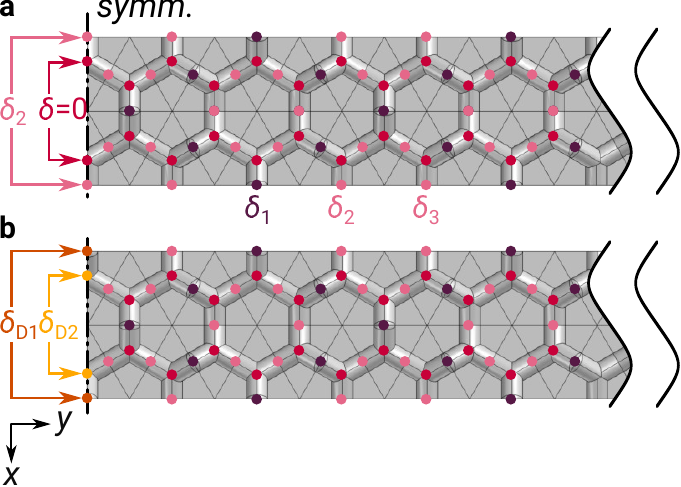}
\caption{\label{fig:optimal} The symmetric dislocation interfaces. (a) The dislocation aligned with an $x$-parallel groove with distortion strength $\delta_2$. The edge states on such dislocation interface are gapped. (b) A modified dislocation interface. The groove depths at the two nodes along the interface are optimized as $\delta_{D1}=0.4$ and $\delta_{D2}=-0.31$ so that the counter-propagating edge states are gapless and fully decoupled.
}
\end{figure}

To create a dislocation that is also a plane of mirror symmetry, we select the configuration with $\phi=0$ so that the bulk lattice possesses $C_{6v}$ symmetry and its mirror image with respect to any plane normal to the $y$-axis still possesses the same bulk pattern. The dislocation is then aligned with an $x$-parallel groove (with distortion strength $\delta_2$) to retain the overall honeycomb-like pattern across the dislocation, as shown in Fig.~\ref{fig:optimal} (a). However, the two counter-propagating edge states are strongly coupled, exhibit mixed pseudospin polarization, and are gapped.

Hence, we allow the groove nodes along the dislocation to take different depths, as illustrated in Fig.~\ref{fig:optimal} (b). We sweep through different values of the distortion strength at the dislocation $(\delta_{D1}, \delta_{D2})$ and find a (non-unique) set of values $(0.4,-0.31)$ such that the edge states' gap closes without mixing the pseudospins, as shown in Fig.~4 in the main text.
\section{Fabrication and experiment setup}
\label{app:exp}
An $18''\times 18''\times 1/4''$ Mic6$^{\mbox{\textregistered}}$ ~ aluminum cast plate was used to fabricate the phononic plate. The cast aluminum alloy plate provides better flatness and lower residual stress over cold-rolled aluminum sheets, which ultimately reduces warpage after machining. The honeycomb groove patterns on both sides were engraved by a ball-end mill programmed to cut through the defined routes on a CNC mill.
%The pattern is a lattice with a Z-shape symmetric dislocation described in the previous section separating two domains of the same bulk lattice. 
A maximum warping deflection of $0.017''$ at the plate center was found after machining one side of the plate. This resulted in overcutting in the groove depth on the other side, which ultimately produced a slight asymmetry with respect to the mid-plane. A 2-inch margin on the boundary of the plate was left untreated (i.e. not machined) so to apply a 3M\texttrademark ~Damping Foil 2552 to reduce unwanted wave reflection from the edges during the test. Two piezoelectric ceramic lead zirconate titanate (PZT-5A) plates were used as actuators. The electric signal was amplified by a Trek$^{\mbox{\textregistered}}$ ~PZD350A M/S amplifier. A Polytec PSV-500 scanning laser-Doppler vibrometer system was used to measure the response of the phononic plate.

\begin{figure*}[ht]
\includegraphics[width=0.8\textwidth]{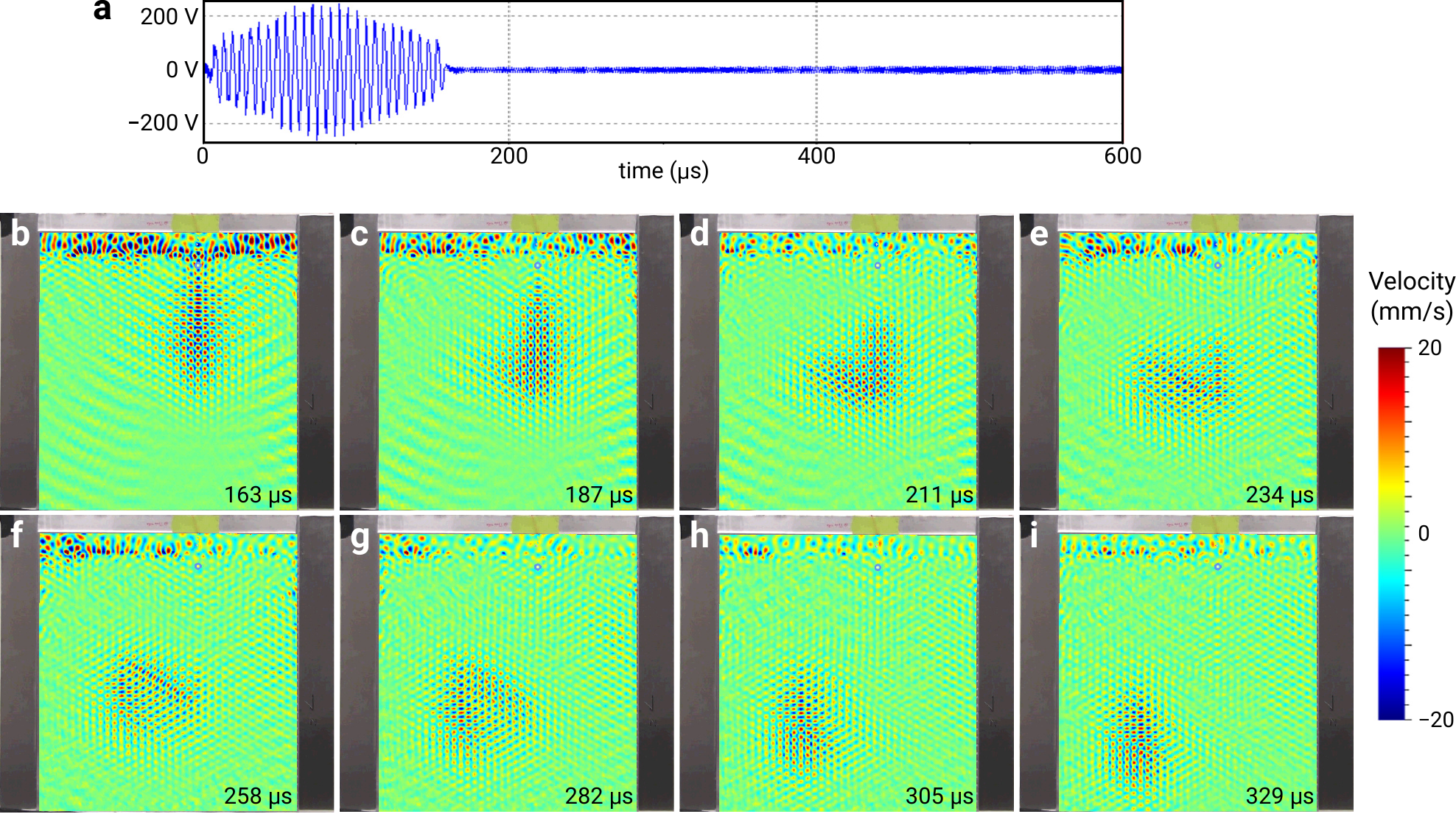}%
\caption{\label{fig:time}
The response of the phononic plate to a transient excitation.
(a) The wave packet used as input signal. (b-i) Instantaneous velocity profiles at selected time frames. The wave packet can travel undisturbed through the two acute corners without observable reflections.
}
\end{figure*}

We also conducted a time-transient measurement following an input wave packet signal centered at 172 kHz as shown in Fig.~\ref{fig:time} (a). Instantaneous velocity profiles at selected time frames are shown in Fig.~\ref{fig:time} (b-i). Results evidently show that the wave packet propagates through the two acute corners without observable reflections. The faster wave with longer wavelength (visible in early time frames (b-d)) is the symmetric bulk mode triggered during the ramp up phase of the transient.
A full video of the transient response is also provided in the online Supporting Information \cite{Supp}.

The results shown in Fig.~\ref{fig:exp} (c,d) in the main text were obtained using a white noise input signal, and the frequency response was obtained by measuring the velocity spectrum normalized by the input voltage spectrum. In the transient measurement, a Kaiser window was used to generate the wave packet. However, due to the capacitive load of the PZT plate, the signal was moderately distorted. The signal shown in Fig.~\ref{fig:time} (a) is the monitored output signal of the amplifier, that is the actual signal applied to the actuator.

%\nocite{*}

\bibliography{REF_v3}% Produces the bibliography via BibTeX.

\end{document}